\documentclass[12pt,aps,pra,amsmath,amssymb,bibnotes]{revtex4}

\usepackage{color,epsfig,rotating,graphicx,subfigure}

\usepackage[latin1]{inputenc}
\usepackage[english]{babel}

\usepackage{dsfont}
\renewcommand{\Im}{{\rm Im}}

\newcommand{\ri}{{\rm i}}
\newcommand{\re}{{\rm e}}
\newcommand{\rd}{{\rm d}}

\newcommand{\alphamat}{\underline{\underline{\alpha}}}

\begin{document}
\title{Magneto-thermoplasmonics: from theory to applications}

\date{\today}

\author{Annika Ott$^1$, Riccardo Messina$^2$, Philippe Ben-Abdallah$^2$, Svend-Age Biehs$^{1}$}
\affiliation{$^1$ Institut f\"{u}r Physik, Carl von Ossietzky Universit\"{a}t, D-26111 Oldenburg, Germany}
\affiliation{$^2$ Laboratoire Charles Fabry,UMR 8501, Institut d'Optique, CNRS, Universit\'{e} Paris-Sud 11,
2, Avenue Augustin Fresnel, 91127 Palaiseau Cedex, France}

\maketitle

\newpage

\noindent
{\bf Abstract:} We review recent theoretical developments on the nanoscale radiative heat transfer in magneto-optical many-particle systems. We discuss in detail the circular heat flux, the giant magneto-resistance effect, the persistent heat current, and the thermal Hall effect for light in such systems within the framework of fluctuational electrodynamics, using the dipolar approximation. We show that the directionality of heat flux in such systems can in principle be understood by analyzing the competing contributions to the heat exchange of the magnetic-field-dependent dipolar resonances of quantum numbers $m = +1$ and $m = -1$. Some potential applications of these effects to thermal and magnetic sensing are also briefly discussed.


\newpage

\section{Introduction}

In the last decade many works devoted to the possible passive control of thermal radiation at the nanoscale have been published. However, works dealing with an active control of nanoscale heat radiation are rather rare. For example, an active switching and tunability of the heat flux was recently discussed by means of an external biasing of graphene layers~\cite{YangWang2017} or ferroelectric materials~\cite{HuangEtAl2014}. A relatively recent trend, the so-called thermotronics~\cite{PBASAB2015}, aims to introduce building blocks like diodes~\cite{PBASAB2013,vanZwol1,NefzaouiEtAl2014}, memories~\cite{Kubytskyi2014,DyakovEtAl2015}, and transistors~\cite{PBASAB2014,OrdonezEtAl2016} for radiative heat flux by employing the properties of phase-change materials like VO$_2$, for instance. Such devices also allow an efficient active control of nanoscale heat fluxes by external heating or cooling and even pave the way to the opportunity of realizing Boolean operations with thermal radiation~\cite{PBASAB2016}. The first experimental proofs of the working principles of the diode and the memory have been realized both in the far- and near-field regime~\cite{ItoEtAl2014,ItoEtAl2016,FiorinoEtAl2018}.

All these works have in common that they show potential techniques to control the magnitude of nanoscale heat flux. Recently, it was shown that in magneto-optical systems the application of an external magnetic field allows for an active control not only of the magnitude of the heat flux, but also of its direction. This directional control is a direct consequence of the non-reciprocal response in such systems. As a result of this behavior, interesting effects can be expected like a circular heat flux carrying also angular momentum~\cite{OttEtAl2018}, a persistent heat current in local and global thermal equilibrium~\cite{Zhu2016}, a giant magneto-resistance effect~\cite{Latella2017,Cuevas}, and a Hall or Righi-Leduc effect for heat radiation~\cite{Ben-Abdallah2016}. Such effects represent potential paths to an ultrafast modulation of the magnitude and direction of heat fluxes at the nanoscale and might be exploited in several applicative domains, including e.g magnetic and thermal sensing.

In this paper, we discuss all these magneto-optical effects for many-particle system within the framework of Rytov's fluctuational electrodynamics, using the dipolar approximation. The general expressions for the heat flux between spherical nanoparticles particles have been derived first in Ref.~\cite{PBAEtAl2011} and then they have been extended to take the radiation correction and the thermal background radiation into account in Ref.~\cite{MessinaEtAl2013}. Finally, this theory has been extended to treat the magnetic polarizabilities and the coupling between the electric and magnetic polarizabilites~\cite{DongEtAl2017}, and to treat anisotropic and non-reciprocal particles~\cite{Saaskilathi2014,Nikbakht2014,EkerothEtAl2017}. This approach has then be used to study a three-body amplification~\cite{PBAEtAl2011}, superdiffusion in plasmonic networks~\cite{PBAEtAl2013}, nanoscale heat transfer between gold arrays~\cite{PHanEtAl2013}, anisotropy effects in many-body configurations~\cite{Nikbakht2014}, heat fluxes in fractal structures~\cite{Nikbakht2017}, to model larger objects within the discrete
dipole approximation~\cite{Edalatpour2014,EkerothEtAl2017}. Furthermore the many-particle theory has been used to study the limitations of the kinetic approach based on the Boltzmann equation~\cite{PBA2008,Jose2015,Jose2018} by using the exact theory to describe the heat flux in a nanoparticle chain~\cite{KathmannEtAl2018}. The impact of the surface mode on a substrate has also been discussed~\cite{Saaskilathi2014,DongEtAl2018,MessinaEtAl2018}.

In the following we start by introducing the basics of the many-body theory based on fluctuational electrodynamics in Sec.~\ref{Sec_Theory}, then we introduce the non-reciprocal material properties of indium antimonide (InSb) in the presence of an external magnetic field in Sec.~\ref{Sec_InSb}. We then discuss the above mentioned magnetic effects in Secs.~\ref{Sec_Circular}--\ref{Sec_Hall} before proposing some potential applications in Sec.~\ref{Sec_Sensing}. Finally, in Sec.~\ref{Sec_Summary} we drive some conclusive remarks.

\section{Many-body theory}\label{Sec_Theory}

We consider an assembly of $N$ identical nanoparticles in local thermal equilibrium at temperatures $T_i$ ($i = 1, \ldots, N$). We assume that the radii of all the nanoparticles are small compared to the other relevant lengthscales in the system, i.e. all the interparticle distances and the wavelengths playing a relevant role in the heat-flux spectrum. This assumption allows us to treat the system within the dipole approximation, according to which each particle is effectively described in terms of a fluctuating (electric and/or magnetic) dipole moments $\mathbf{p}^{\rm (fl)}_i$, generating an electric and magnetic field given by
\begin{equation}
\begin{split}
 \mathbf{E}^{\rm (fl)}(\mathbf{r}) = \mu_0\omega^2\sum_{i}\mathds{G}^{\rm E}(\mathbf{r},\mathbf{r}_i)\mathbf{p}^{\rm (fl)}_i, \\
 \mathbf{H}^{\rm (fl)}(\mathbf{r}) = \mu_0\omega^2\sum_{i}\mathds{G}^{\rm H}(\mathbf{r},\mathbf{r}_i)\mathbf{p}^{\rm (fl)}_i,
\end{split}
\end{equation}
expressed by means of the electric and magnetic Green tensors $\mathds{G}^{\rm E/H}$. In free space, these are known analytically and read
\begin{equation}
\begin{split}
 \mathds{G}^{\rm E}(\mathbf{r},\mathbf{r}_i,\omega) &= \frac{e^{{\rm i} k_0 d}}{4\pi d}\left[a\mathds{1}+b\mathbf{e}_d\otimes\mathbf{e}_d\right], \\
 \mathds{G}^{\rm H}(\mathbf{r},\mathbf{r}_i,\omega) &= \frac{e^{{\rm i}k_0d}}{4\pi d}l\frac{1}{\mu_0c}(\mathbf{e}_\phi\otimes\mathbf{e}_\theta-\mathbf{e}_\theta\otimes\mathbf{e}_\phi),
\end{split}
\end{equation} 
with 
\begin{equation}
\begin{split}
 a &= 1+\frac{{\rm i}k_0 d -1}{k_0^2 d},\\
 b &= \frac{3-3{\rm i}k_0 d -k_0^2 d^2}{k_0^2 d},\\
 l &= 1+\frac{{\rm i}}{k_0d},
\end{split}
\end{equation} 
where we have introduced the wavenumber in vacuum $k_0 = \omega/c$, the vacuum permeability $\mu_0$, the relative distance $d = |\mathbf{r} - \mathbf{r}_i|$ between the position $\mathbf{r}_i$ of particle $i$ and the observation point $\mathbf{r}$, and the unit vector $\mathbf{e}_d = (\mathbf{r} - \mathbf{r}_i)/d$. The unit vectors $\mathbf{e}_\phi$ and $\mathbf{e}_\theta$ are the standard azimuthal and polar unit vectors with respect to a coordinate system with its origin in $\mathbf{r}_i$.

In order to take the mutual interaction of the nanoparticles into account we have to consider the total field $\mathbf{E} = \mathbf{E}^{\rm (fl)} + \mathbf{E}^{\rm (ind)}$ which is the sum of the thermal fields generated by the fluctuating dipole moments $\mathbf{p}_i^{\rm (fl)}$ and the field 
\begin{equation}
  \mathbf{E}^{\rm (ind)}(\mathbf{r}) = \mu_0\omega^2 \sum_{i}\mathds{G}^{\rm E}(\mathbf{r},\mathbf{r}_j)\mathbf{p}_i^{\rm (ind)},\label{Eind}
\end{equation}
generated by the induced dipole moments 
\begin{equation}
 \mathbf{p}_i^{\rm (ind)} = \epsilon_0\underline{\underline{\alpha}}\mathbf{E}^{\rm (ind)}(\mathbf{r}_i). \label{ind}
\end{equation}
The strength of the induced dipole moments is determined by the polarizability tensor $\underline{\underline{\alpha}}$ which will be specified later. Of course, similar expressions hold for the magnetic field. 

These expressions allow us to determine the correlation functions of the fluctuating electromagnetic field in the $N$-particle configuration. To this aim, we apply the fluctuation-dissipation theorem of the second kind~\cite{Callen} to the fluctuating dipole moments 
\begin{equation}\label{FDT}
  \langle p^{\rm (fl)}_i (\omega) \otimes {p^{\rm (fl)*}_i}(\omega')\rangle = \frac{2\epsilon_0}{\omega}\Theta(\omega,T_i)\frac{1}{2{\rm i}}(\underline{\underline{\rm \alpha}} - \underline{\underline{\rm \alpha}}^\dagger) \delta(\omega - \omega'),
\end{equation} 
where the brackets denote an ensemble average. Note that here we neglect the radiation correction~\cite{Cuevas} which is negligibly small in the configurations described here.

After some extensive algebra we obtain the following expression of the spectral mean Poynting vector
\begin{equation}
 \begin{split}
  \langle S_{\omega,\alpha}\rangle &= 2{\rm Re} \langle\mathbf{E}\times\mathbf{H}^*\rangle_\alpha \\ 
                   &= 4k_0^2\omega\mu_0\epsilon_{\alpha\beta\gamma} 
                    {\rm Re}\Biggl[\sum_{k=1}^{N}\Theta(\omega,T_k) \Biggl(\sum_{i=1}^{N}\mathds{G}_{0i}^{\rm E}\boldsymbol{T}_{ik}^{-1}\Biggr)_{\beta\delta} \\ &\qquad \Biggl(\frac{\alphamat-\alphamat^\dagger}{2 \rm i}\Biggr)_{\delta\epsilon}\Biggl(\sum_{j=1}^{N}\mathds{G}_{0j}^{\rm H}\boldsymbol{T}_{jk}^{-1}\Biggr)^\dagger_{\epsilon\gamma}\Biggr],
\end{split}
\label{poyntingN}
\end{equation}
where $\epsilon_{\alpha\beta\gamma}$ is the Levi-Civita tensor (using Einstein notation) and the Greek indices stand for the components of vectors and tensors, and $\mathds{G}_{0i}^{\rm E/H} :=\mathds{G}^{\rm E/H}(\mathbf{r},\mathbf{r}_i)$. Furthermore, we have introduced the $ \boldsymbol{T}$ matrix having elements
\begin{equation}
 \boldsymbol{T} _{ij} = \delta_{ij}\mathds{1}-(1-\delta_{ij})k_0^2\alphamat\mathds{G}^{\rm E}_{ij}.
\label{TT}
\end{equation}
These are fully determined by the polarizability and Green tensors, for which we have introduced the notation $\mathds{G}_{ij}^{\rm E/H} :=\mathds{G}^{\rm E/H}(\mathbf{r}_i,\mathbf{r}_j)$. The full mean Poynting vector is as usual
given by the integral expression
\begin{equation}
 \langle \mathbf{S} \rangle = \int_0^\infty \frac{\rd \omega}{2 \pi} \, \langle \mathbf{S}_{\omega}\rangle.
\end{equation}

Similarly, we can derive the expression for the power transferred from particle $j$ to particle $i$ and obtain
\begin{equation}
 \begin{split}
   \langle P_{ij}\rangle &= \biggl\langle\frac{d\mathbf{p}_i(t)}{dt}\cdot\mathbf{E}_{ij}(t)\biggr\rangle \\
              &= 3\int_{0}^{\infty}\frac{{\rm d}\omega}{2\pi}\Theta(\omega,T_j)\mathcal{T}_{ij}(\omega),
 \end{split}
\label{waermestromNij}
\end{equation}
with the transmission coefficient defined as
\begin{equation}
 \mathcal{T}_{ij}(\omega) = \frac{4}{3}{\rm Im Tr}\left[\boldsymbol{T}^{-1}_{ij}\frac{\alphamat-\alphamat^\dagger}{2 \rm i}(\boldsymbol{T}^{-1}_{ij})^\dagger\alphamat^{-1\dagger}\right].
\label{tauik}
\end{equation}
Note that for non-reciprocal permittivities in general $ \mathcal{T}_{ij} \neq \mathcal{T}_{ji}$, so that the net power received by particle $i$ is given by
\begin{equation}
  \langle P_{i}\rangle = \sum_{j \neq i}^{N} 3\int_{0}^{\infty}\frac{{\rm d}\omega}{2\pi}\Big(\Theta(\omega,T_j)\mathcal{T}_{ij}(\omega)-\Theta(\omega,T_i)\mathcal{T}_{ji}(\omega)\Big).
\label{waermenetto}
\end{equation}
This means that in our formalism if the particle $i$ receives (emits) a net power then $ \langle P_{i} \rangle > 0$ ($\langle P_{i} \rangle < 0$). More details on the derivation of $\langle P_{ij}\rangle$ can be found in Ref.~\cite{MessinaEtAl2013}. We stress here that, while in the case of reciprocal materials the direction of heat flux from particle is entirely determined by the temperatures (from hotter to colder particle), Eq.~\eqref{waermenetto} shows that an asymmetry in the transmission coefficients can non-trivally affect the value of the flux and also, in principle, its sign.

\section{Material properties}\label{Sec_InSb}

In the previous section, we have described each nanoparticle in terms of a fluctuating dipole, whose correlation function is connected [by means of the fluctuation-dissipation theorem, see Eq.~\eqref{FDT}] to the polarizability tensor $\alphamat$. This matrix can be in turn expressed in terms of the permittivity tensor $\underline{\underline{\epsilon}}$ as~\cite{LakhtakiaEtAl1991,Albaladejo}
\begin{equation}
\underline{\underline{\alpha}} = 4\pi R^3(\underline{\underline{\epsilon}}-\mathds{1})(\underline{\underline{\epsilon}}+2\mathds{1})^{-1}.
\label{alphaerg}
\end{equation}
This expressions shows that the properties of the permittivity tensor will have a direct impact on the polarizability: in particular, a diagonal (non-diagonal) permittivity tensor will result in a diagonal (non-diagonal) polarizability $\alphamat$.

\begin{equation}
 \underline{\underline{\rm \epsilon}}=\begin{pmatrix} \epsilon_1 & -{\rm i}\epsilon_2 & 0 \\ {\rm i}\epsilon_2 & \epsilon_1 & 0 \\ 0 & 0 & \epsilon_3 \end{pmatrix}.
\label{epsilon}
\end{equation}
Note that the system is non-diagonal and non-reciprocal since $\underline{\underline{\epsilon}} \neq \underline{\underline{\epsilon}}^t$. Therefore, the polarizability matrix shares the same properties and can be written as
\begin{equation}
  \underline{\underline{\alpha}}=\begin{pmatrix} \alpha_1 & \alpha_{12} & 0 \\ \alpha_{21} & \alpha_1 & 0 \\ 0 & 0 & \alpha_3 \end{pmatrix},
\end{equation}
with $\alpha_{12} = -\alpha_{21}$. The appearance of the non-diagonal elements and the non-reciprocity are due to the Lorentz force or, more in general, to the time-reversal symmetry breaking induced by the presence of the external magnetic field.

In the following we focus on the material properties of n-doped InSb, i.e.\ one possible example of material whose optical response can be tuned by means of an external magnetic field. For InSb the components of the permittivity tensor are determined by the phononic and electronic response which can be described by a Drude-Lorentz and Drude model, respectively. Introducing the cyclotron frequency $\omega_c = e B /m^*$, we have~\cite{Palik}
\begin{equation}
\begin{split}
 \epsilon_1 &= \epsilon_\infty\biggl(1+\frac{\omega_{\rm L}^2-\omega_{\rm T}^2}{\omega_{\rm T}^2-\omega^2-{\rm i}\Gamma\omega}  +\frac{\omega_{\rm p}^2(\omega+{\rm i}\gamma)}{\omega[\omega_{\rm c}^2-(\omega+{\rm i}\gamma)^2]} \biggr), \\
 \epsilon_2 &= \frac{\epsilon_\infty\omega_{\rm p}^2\omega_{\rm c}}{\omega[(\omega+{\rm i}\gamma)^2-\omega_{\rm c}^2]},\\ 
 \epsilon_3 &= \epsilon_\infty\left(1+\frac{\omega_{\rm L}^2-\omega_{\rm T}^2}{\omega_{\rm T}^2-\omega^2-{\rm i}\Gamma\omega}-\frac{\omega_{\rm p}^2}{\omega(\omega+{\rm i}\gamma)} \right).
\end{split}
\end{equation}

The effects which we are going to discuss are very sensitive to the chosen material parameters. Therefore, in the following we use two different sets of material properties which correspond to two different doping levels of InSb. On the one hand, we use as parameter-set A the data from Ref.~\cite{Ben-Abdallah2016}, namely $n = 1.36\times10^{19}$\,cm$^{-3}$, $m^* = 7.29\times10^{-32}$\,kg, $\omega_{\rm p} = \sqrt{\frac{ne^2}{m^*\epsilon_0\epsilon_\infty}} = 1.86\times10^{14}$\,rad/s, and $\gamma = 10^{12}$\,rad/s. On the other hand, as parameter-set B we take the values from Ref.~\cite{Zhu2016}  $n = 1.07\times10^{17}$\,cm$^{-3}$, $m^* = 1.99\times10^{-32}$\,kg, $\omega_{\rm p} = \sqrt{\frac{ne^2}{m^*\epsilon_0\epsilon_\infty}} = 3.15\times10^{13}$\,rad/s, and $\gamma = 3.39\times10^{12}$\,rad/s. For both parameter sets the phononic response is described by $\epsilon_\infty = 15.7$, $\omega_{\rm L} = 3.62\times 10^{13}\,{\rm rad/s}$, $\omega_{\rm T} = 3.39\times10^{13}\,{\rm rad/s}$, and $\Gamma = 5,65\times10^{11}$\,rad/s. Parameter set B is used in Sec.~\ref{Sec_Circular}, whereas parameter set A is used in Secs.~\ref{Sec_Resistance}, \ref{Sec_Persistent} and \ref{Sec_Hall}.

\section{Circular heat flux}\label{Sec_Circular}

\begin{figure}
	\centering
	\includegraphics[width=0.45\textwidth]{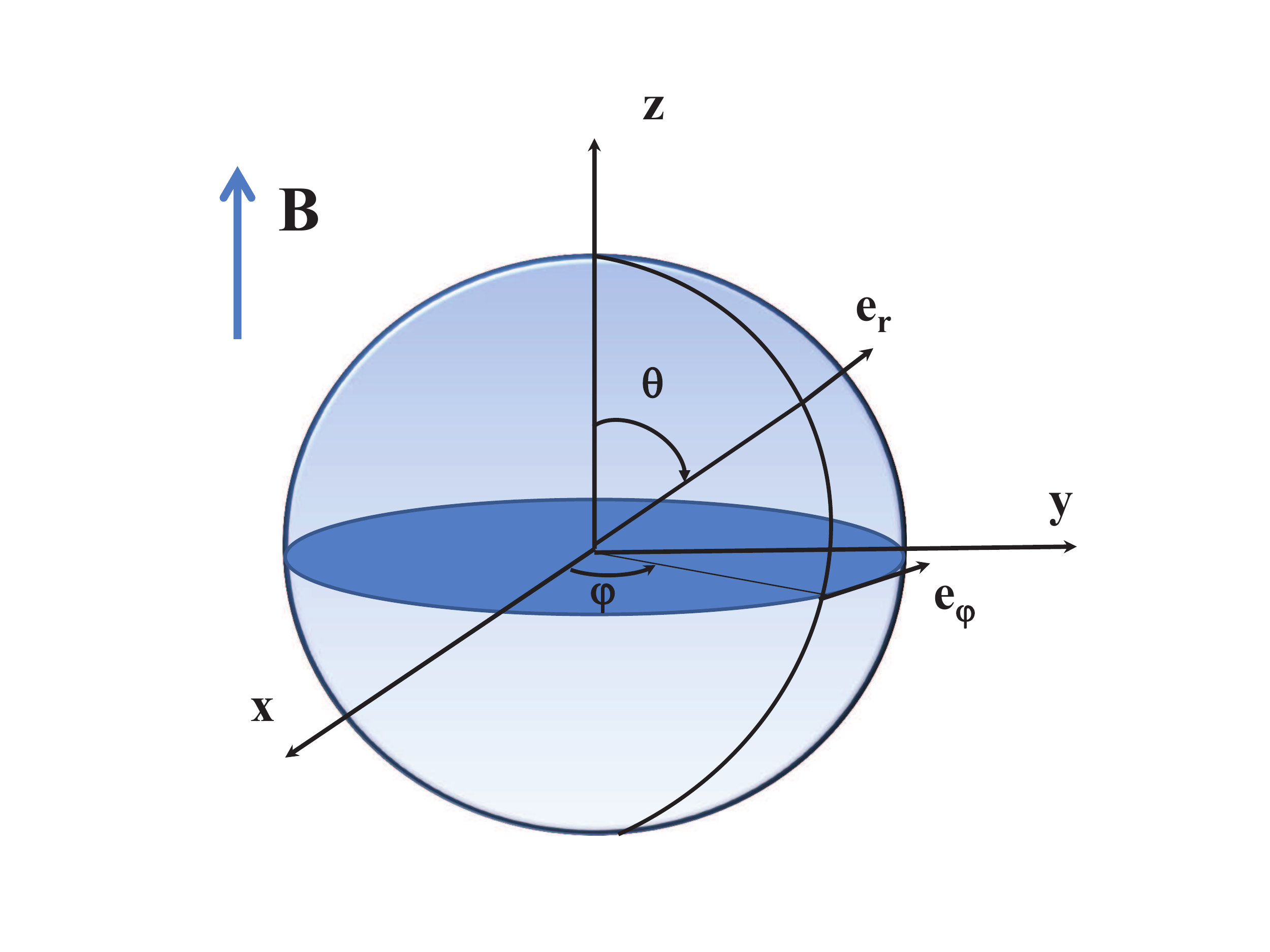}
	\caption{Sketch of a magneto-optical nanoparticle and associated coordinate system.}
	\label{MOparticle}
\end{figure}

We first analyze the radiative behavior of a single particle (as sketched in Fig.~\ref{MOparticle}) with temperature $T_1 = T_p$ placed in the origin of our coordinate system. These results will serve as the basis for the discussion of heat fluxes and currents in many-body systems. In this case the only relevant physical quantity is the Poynting vector, with the advange that it can be evaluated analytically. Expressing our result in spherical coordinates we find~\cite{OttEtAl2018}
\begin{equation}
 \langle \mathbf{S}_{\omega}\rangle = S_{r,\omega}\mathbf{e}_r+S_{\theta,\omega}\mathbf{e}_{\theta}+S_{\phi,\omega}\mathbf{e}_{\phi},
\label{poynting}
\end{equation} 
with components
\begin{equation}\label{Somega}
\begin{split}
 S_{r,\omega} &= \frac{\Theta(\omega,T_{\rm p})k_0^3}{4\pi^2r^2}\Big(\alpha_{11}'' [1 + \cos^2(\theta)]+\alpha_{33}''\sin^2(\theta)\Big),\\ 
 S_{\phi,\omega} &= \frac{\Theta(\omega,T_{\rm p}) k_0^3}{4\pi^2r^2}2\alpha_{12}'\left(\frac{1}{k_0r}+\frac{1}{k_0^3r^3}\right)\sin(\theta),\\
 S_{\theta,\omega} &= 0.
 \end{split}
\end{equation}
For evident symmetry reasons, the Poynting vector has only a radial and an azimuthal component, but no polar one. In particular, the azimuthal component solely depends on the non-diagonal element $\alpha_{12}$ and is therefore a direct consequence of the Lorentz force on the electrons in the nanoparticle. For the case $B = 0$ the Poynting vector is purely radial and $\alpha_{11}'' = \alpha_{33}'' \equiv \alpha''$ so that we retrieve the well-known Mie-Planck  formula~\cite{BohrenHuffman}
\begin{equation}
 \langle \mathbf{S}_{\omega}\rangle = \frac{\Theta(\omega,T_p)k_0^3}{2\pi^2r^2} \alpha'' \mathbf{e}_r
\end{equation}
with the typical $1/r^2$ dependence which guarantees that the flux through a spherical surface around the nanoparticle is constant for any choice of the radius of that sphere, so that the total emitted energy per unit time is constant.

\begin{figure}
	\centering
	 \includegraphics[width=0.45\textwidth]{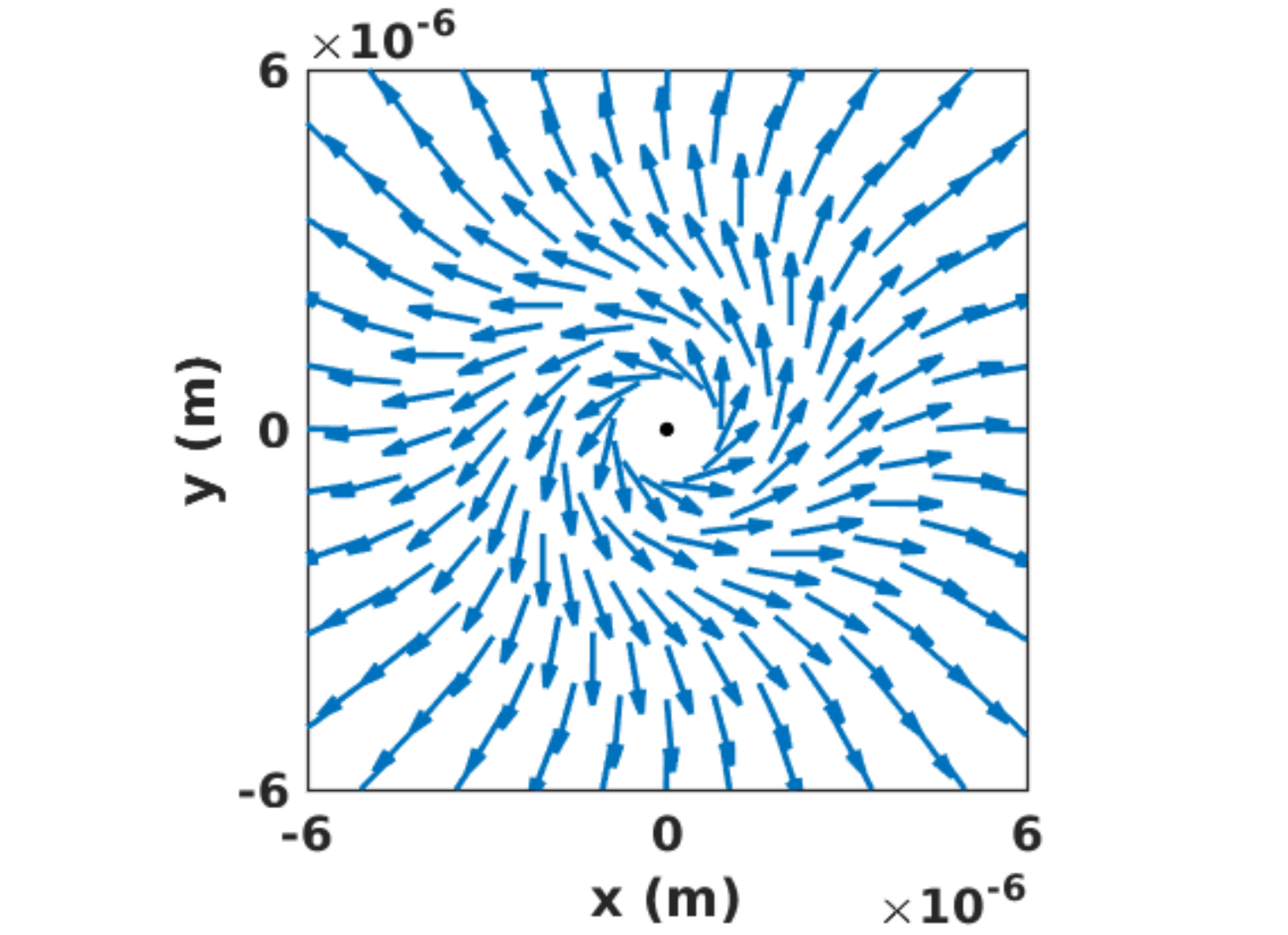}
	\caption{Mean Poynting vector around a InSb nanoparticle with radius $R=100\,$nm and at temperature $T_p=300\,{\rm K}$, immersed in a vacuum background with temperature $T_b = 0\,{\rm K}$. A magnetic field of 1\,T in the positive $z$ direction is applied. The plotted vortex-like vector field in the $x$-$y$ plane shows the transition between the near-field and far-field regime around a distance of about 6\,$\mu$m.}
	\label{circular}
\end{figure}

In the presence of a non-vanishing external magnetic field $\mathbf{B} = B \mathbf{e}_z$, the Poyinting vector has an azimuthal component and therefore the flux lines (i.e. the flow lines of the Poynting vector field) around the nanoparticle are circular with respect to the $z$ axis. As it can be seen in Fig.~\ref{circular}, in the near-field regime the full heat flux in the plane perpendicular to the $z$ axis is circulating in counterclockwise direction, whereas in the far-field regime it is circulating in clockwise direction. To have a basic understanding of the mechanism behind this circular heat flux, we study the dipole resonances with angular momentum $l = 1$. These resonances are determined by the poles of the polarizability, i.e.\ by the conditions
\begin{equation}
\begin{split}
 \epsilon_1 + 2 &= \mp \epsilon_2 , \qquad (\text{for } m = \pm 1), \\
 \epsilon_3 + 2 &= 0, \hspace{1.4cm} (\text{for } m = 0).
\end{split}
\end{equation}
For $B = 0$ the three resonances with magnetic quantum number $m = 0, \pm 1$ are degenerate, because in this case $\epsilon_2 = 0$ and $\epsilon_1 = \epsilon_3$. When applying a magnetic field $B \neq 0$ the degeneracy is lifted due to the Lorentz force acting on the oscillating electrons. Neglecting for a moment the phononic contribution in the permittivity and setting $\gamma = 0$, we find the analytical expressions
\begin{equation}
\begin{split}
  \omega_{m = \mp1} &= \sqrt{\left(\frac{\epsilon_\infty\omega_{\rm p}^2}{\epsilon_\infty+2}+\frac{\omega_{\rm c}^2}{4}\right)} \pm \frac{\omega_{\rm c}}{2}, \\
  \omega_{m = 0} &= \sqrt{\frac{\epsilon_\infty\omega_{\rm p}^2}{\epsilon_\infty+2}},
\end{split}\end{equation}
for the dipolar resonances. It is apparent from these expressions that the resonance with magnetic quantum number $m = 0$ is unaffected by the presence of the magnetic field. On the other hand, the two resonances with $m = \pm1$ split in frequency, the size of the splitting being determined by the cyclotron frequency $\omega_c$. This trend can of course also be observed when including the damping of the electrons and the response of the optical phonons in the permittivity, which are not affected by the magnetic field.

\begin{figure}
	\centering
    \includegraphics[width=0.45\textwidth]{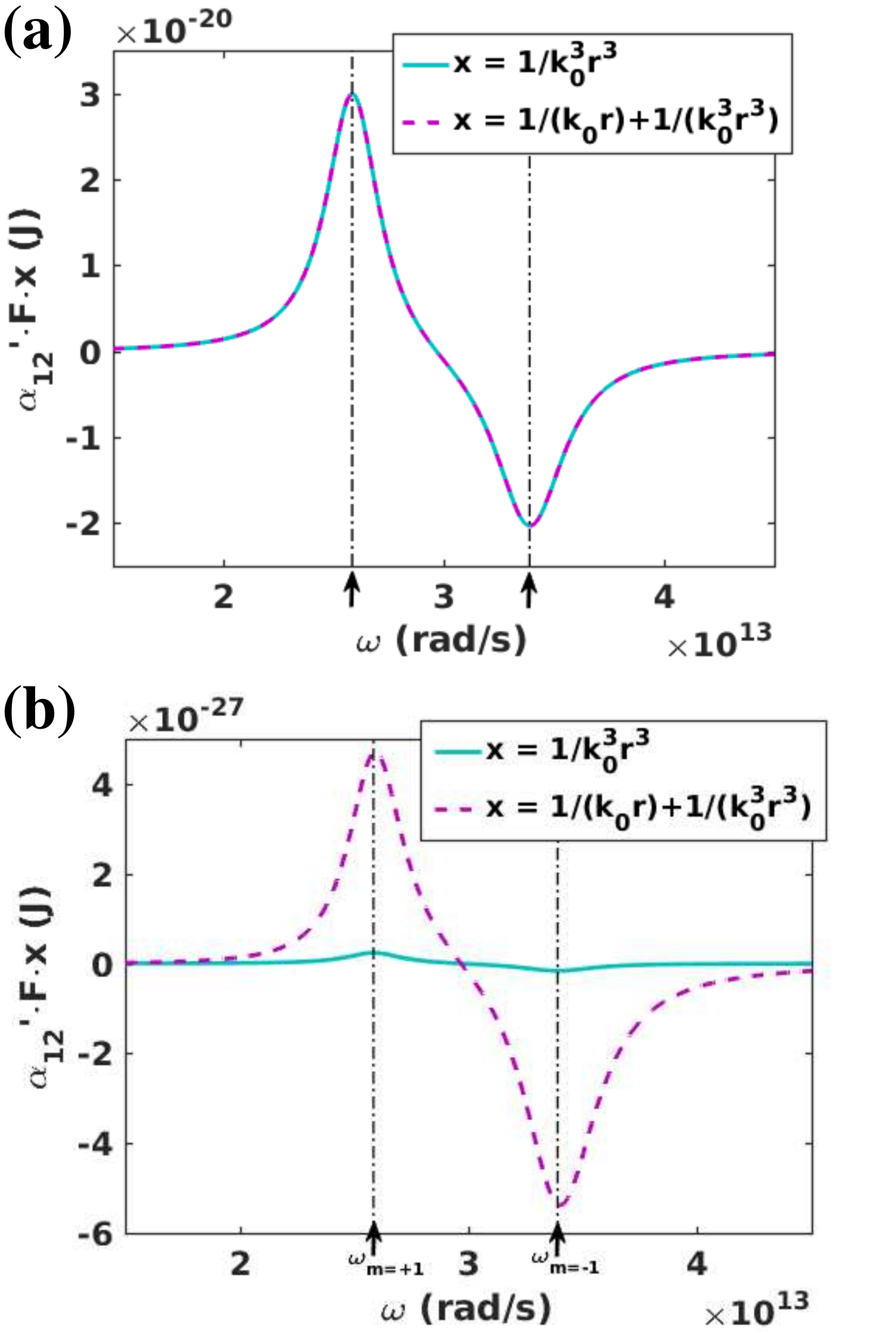}
	\caption{Plot of $\alpha_{12}' \Theta(\omega,T_p) k_0^3$ weighted by the factors $x = 1/(k_0 r)^3$ (near-field regime) and $x = 1 / k_0 r$ (far-field regime), considering only the electric response (neglecting the phononic contribution to the permittity), for a magnetic field of 1\,T at temperature $T_p=300\,$K, $T_b=0\,$K and for $R = 100\,{\rm nm}$. In (a) we have chosen a distance of $r = 100\,{\rm nm}$ and in (b) $r = 50\,\mu {\rm m}$. In the near field the low-frequency mode of the $m = +1$ resonance is dominating, whereas in the far-field regime the $m = -1$ mode dominates the spectral heat flux, so that the direction of the heat flux is changed.}
	\label{alpha12}
\end{figure}

It is now important to note that the spectral Poynting vector at the resonance with $m = +1$ ($m = -1$) is counterclockwise (clockwise) whereas for $m = 0$ the Poynting vector is purely radial. Due to the splitting in frequency, the two resonances with $m = \pm 1$ can contribute differently to the heat flux. More precisely, in the near-field regime with $k_0 r \ll 1$ the azimuthal component of $S_{\phi,\omega}$ in Eq.~\eqref{Somega} is weighted by the factor $1/(k_0 r)^3$, so that in this case the low-frequency resonance with $m = +1$ will dominate and therefore also give the leading contribution to the frequency-integrated heat flux. In the far-field regime, defined by $k_0 r \gg 1$, the azimuthal component of $S_{\phi,\omega}$ in Eq.~\eqref{Somega} is weighted by the factor $1/(k_0 r)$. As shown in Fig.~\ref{alpha12}, in this case the high-frequency resonance $m = -1$ dominates the heat flux. Hence, the transition from the counterclockwise heat flux in the near-field regime to the clockwise one in the far-field regime as observed in Fig.~\ref{circular} can be understood by the fact that in the near-field regime the heat flux is dominated by the resonance with magnetic quantum number $m = +1$ and in the far-field regime by the other resonance with $m = -1$.

Of course, due to the circular heat flux the fluctuating electromagnetic field also carries an angular momentum which can be split into an orbital angular momentum and a spin angular momentum. We have discussed in detail this property in Ref.~\cite{OttEtAl2018} using the definitions of the spin and angular momentum introduced by Bliokh and Nori~\cite{Bliokh1,Bliokh}. However, here we want to focus only on the directionality of the heat flux, and we will see that the observed behavior can be helpful in the interpretation of the directionality of the heat transfer in more complex configurations. Before we discuss such configurations we want to emphasize that we can already for a single particle find the analogue of the persistent heat current found by Zhu and Fan~\cite{Zhu2016}, which will also be discussed later. To this end, we assume that the particle is immersed in an environment populated by thermal photons at temperature $T_b \neq 0$. As detailed in Ref.~\cite{OttEtAl2018} in this configuration the total spectral Poynting vector reads
\begin{equation}
  \mathbf{S}_\omega^{\rm tot} =  \mathbf{S}_\omega(T_p) - \mathbf{S}_\omega(T_b) + \mathbf{S}_\omega^{\rm pers} (T_b),
\end{equation}
with $\mathbf{S}_\omega(T)$ from Eq.~(\ref{poynting}) and
\begin{equation}
  \mathbf{S}_\omega^{\rm pers} = \frac{k_0^3 \Theta(\omega,T_b)}{4\pi^2r^2} \sin(\theta)\Im\biggl( l (a + b) \frac{\alpha_{12} - \alpha_{21}}{2} \re^{2 \ri k r} \biggr) \mathbf{e}_\varphi.
\end{equation}
As a consequence, we deduce that at global thermal equilibrium the total heat flux is not zero, but there is a {\itshape persistent heat flux} $\mathbf{S}_\omega^{\rm pers}$ which is purely azimuthal and has the same circular properties discussed above for $\mathbf{S}_\omega(T)$ [see Eq.~\eqref{poynting}]. As manifest from the analytical expression, this effect only exists for non-reciprocal materials with $\alphamat \neq \alphamat^t$. Note that the heat flux through a spherical surface including the nanoparticle is zero, as well as the one through any infinite plane so that this persistent heat flux does not result in a real thermal emission. 

\section{Giant magneto-resistance}\label{Sec_Resistance}

\begin{figure}
	\centering
	\includegraphics[width=0.45\textwidth]{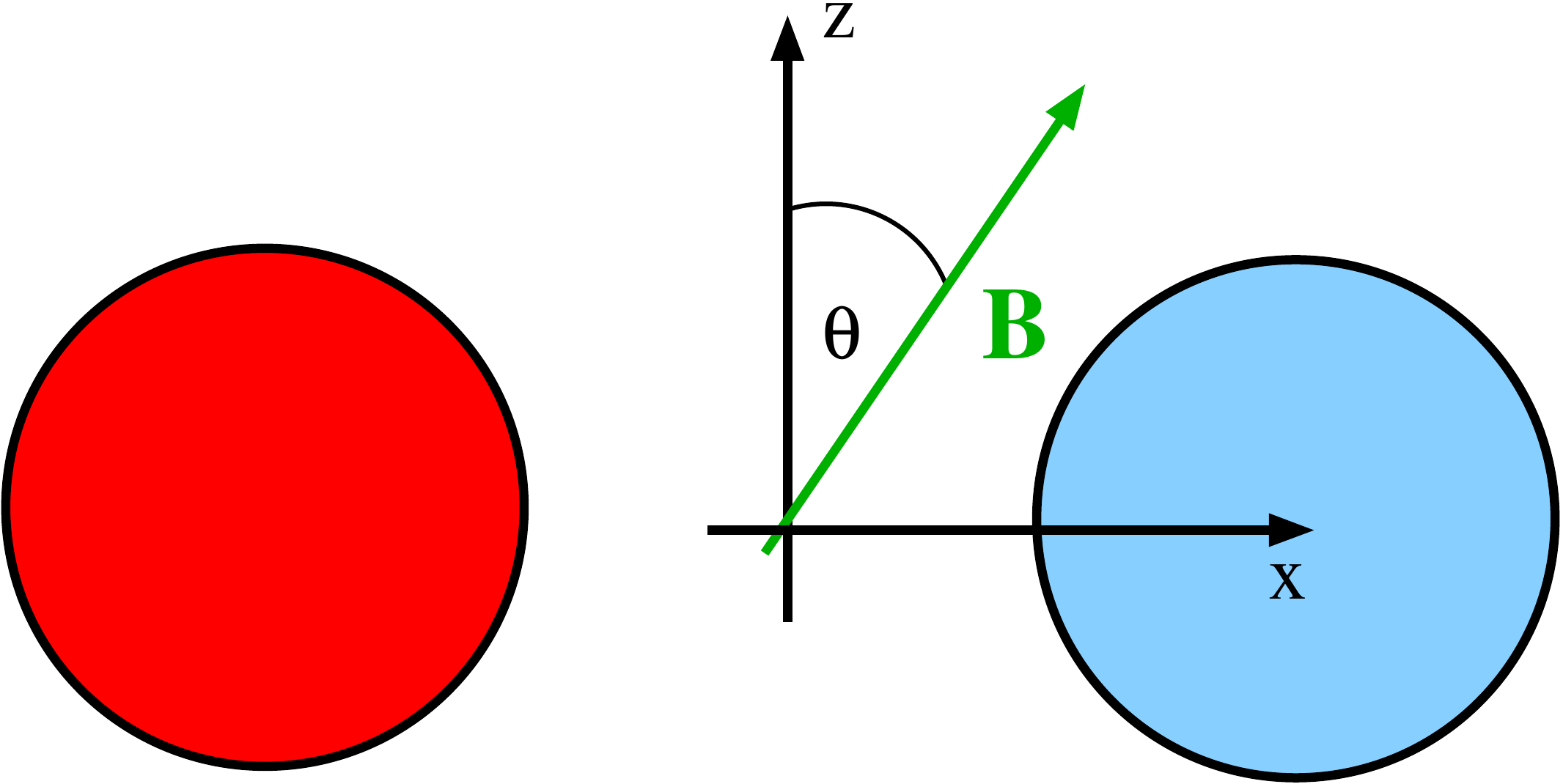}
	\caption{Sketch of two magneto-optical particles in the presence of a magnetic field $\mathbf{B}$ forming an angle $\theta$ with the line connecting the centers of the two particles.}
	\label{MOparticleResist}
\end{figure}

As a next step, we want to study the impact of the presence of the magnetic field on the heat transfer between two nanoparticles with temperatures $T_1 \neq T_2$. In particular, we first study how a change of the magnitude of the magnetic field alters the heat transfer by assuming that the magnetic field is perpendicular to the axis connecting the two particles. Referring to the scheme given in Fig.~\ref{MOparticleResist}, we take here $\theta = 0$. To this end we can use our general expression for $\langle P_{ij} \rangle$ or $\langle P_{i} \rangle$ in Eqs.\eqref{waermestromNij} and \eqref{waermenetto}, respectively, with $i,j = 1,2$. Because of the symmetry of this two-particle configuration, we have $\mathcal{T}_{12} = \mathcal{T}_{21}$, so that 
\begin{equation}
\begin{split}
  \langle P_{1}\rangle &= 3\int_{0}^{\infty}\frac{{\rm d}\omega}{2\pi}\Big(\Theta(\omega,T_2)-\Theta(\omega,T_1)\Big)\mathcal{T}_{12}(\omega) \\
            &= - \langle P_{2}\rangle.
\end{split}
\end{equation}
From $\mathcal{T}_{12} = \mathcal{T}_{21}$ it is obvious that the non-reciprocity does not play a role in this specific scenario, meaning that it does not induce an asymmetry in the transmission coefficients. In particular, for $T_1 = T_2$ we have $ \langle P_{1}\rangle = \langle P_{2}\rangle = 0$ and the heat flux between the particles fulfills the property $\langle P_{12}\rangle = \langle P_{21}\rangle$. Consequently the persistent circular heat flux around the nanoparticles, which also exists in the two-particle configuration, does not result in a persistent heat transfer between the particles. 

\begin{figure}
	\centering
    \includegraphics[width=0.45\textwidth]{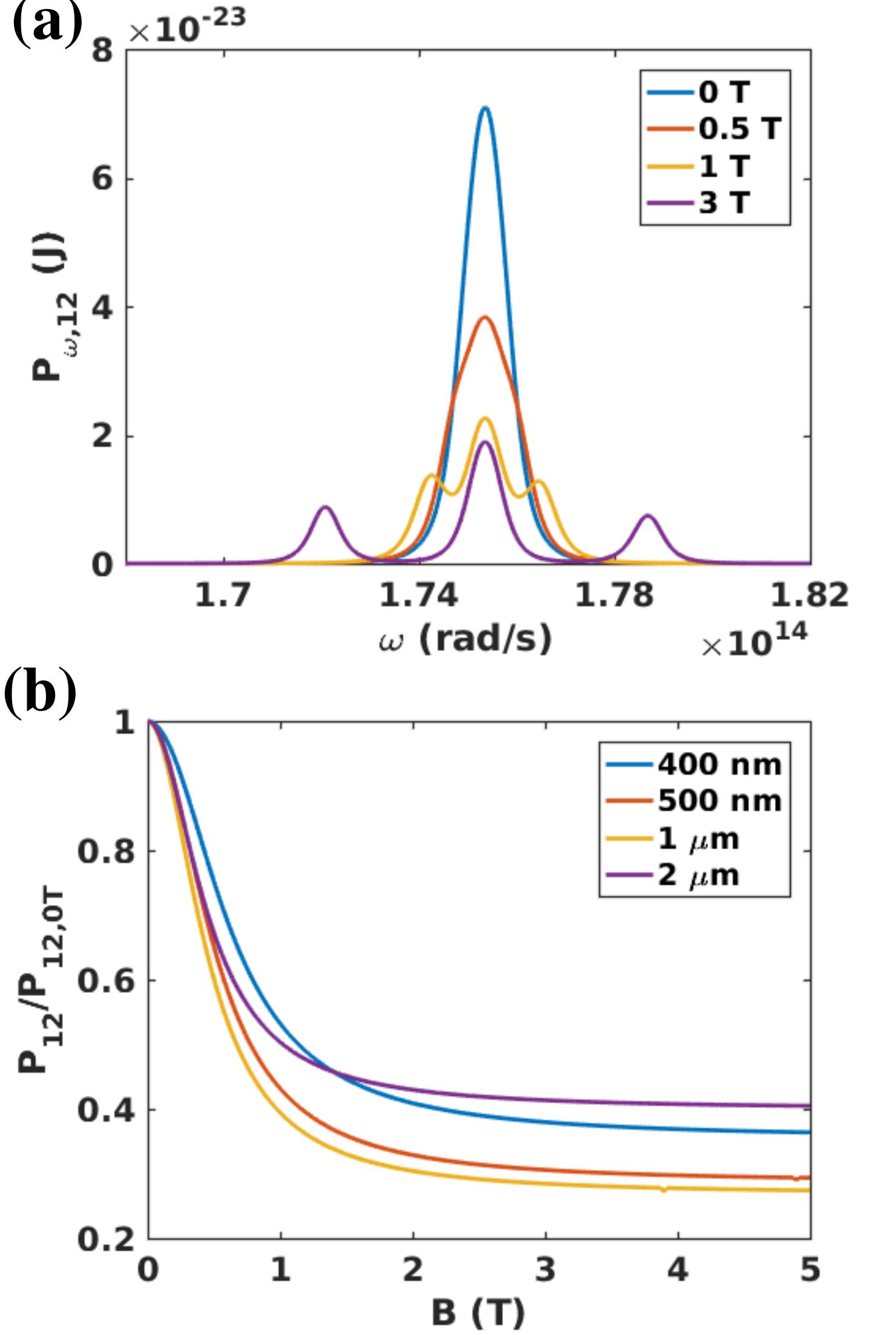}
	\caption{(a) Spectral heat flux $ P_{12,\omega}$ between two InSb nanoparticles of radius $R=100\,$nm at temperatures $T_1=310\,{\rm K}$ and $T_2 = 300\,{\rm K}$ for an interparticle distance of $d = 400\,{\rm nm}$ and different magnetic-field magnitudes. (b) Total transferred power $\langle P_{12} \rangle$ as function of the magnetic field amplitude normalized to $\langle P_{12} (B = 0\,\text{T})\rangle$. }
	\label{SpectralGiantResist}
\end{figure}

However, the splitting of the dipolar resonances has an impact on the magnitude of transferred heat for the case where $T_1 \neq T_2$. This can be nicely seen in Fig.~\ref{SpectralGiantResist}, showing the spectral heat flux $ P_{12,\omega}$ for different magnitudes of the applied magnetic field. For $B = 0$ the three dipolar resonances are degenerate, whereas they start to split when $B$ is increased. Furthermore, the amplitudes of the resonance peaks drop when the field is applied. Both the splitting and the amplitude drop result in an overall drop of the power $|\langle P_{1}\rangle|$ emitted from particle 1 and received by particle 2 when the magnitude of the magnetic field is increased. We observe in Fig.~\ref{SpectralGiantResist} that this power drop in the configuration under scrutiny can achieve values as large as 70\%. This is the giant magneto-resistance effect discussed in detail in Ref.~\cite{Latella2017}.

Instead of changing the magnitude of the heat flux by changing the magnitude of the magnetic field, we can obtain the same result by changing its direction. This property was first highlighted in Ref.~\cite{Cuevas}. To take into account this effect in our calculations, we rotate the permittivity tensor by the angle $\theta$ around the $y$-axis: this is equivalent to a rotation of the magnetic field by the angle $\theta$ around the $y$-axis. In this case the angle-dependent permittivity tensor reads
\begin{equation}
 \underline{\underline{\epsilon}} = {\scriptstyle\begin{pmatrix}  \epsilon_{11} & - \ri \epsilon_2 \cos (\theta) &  \epsilon_{13}\\
                    \ri \epsilon_2 \cos (\theta) & \epsilon_1 & \ri \epsilon_2 \sin(\theta) \\ 
                      \epsilon_{13} & - \ri \epsilon_2 \sin(\theta) &   \epsilon_{33}
                  \end{pmatrix}},
              \end{equation} 
            where
 \begin{equation}
 \begin{split}\epsilon_{11} & =\epsilon_1 \cos^2 (\theta)+ \epsilon_3 \sin^2 (\theta),\\
  \epsilon_{13} & =\frac{1}{2}(\epsilon_1 - \epsilon_3)\sin(2 \theta),\\
   \epsilon_{33} & =\epsilon_1 \sin^2 (\theta)+ \epsilon_3 \cos^2 (\theta).
   \end{split}
\end{equation} 
Since our two particles are aligned along the $x$-axis (see Fig.~\ref{MOparticleResist}), the magnetic field will be perpendicular to the line connecting the particles for $\theta = 0$ and parallel to that line if $\theta = \pi/2$.

\begin{figure}
	\centering
	\includegraphics[width=0.45\textwidth]{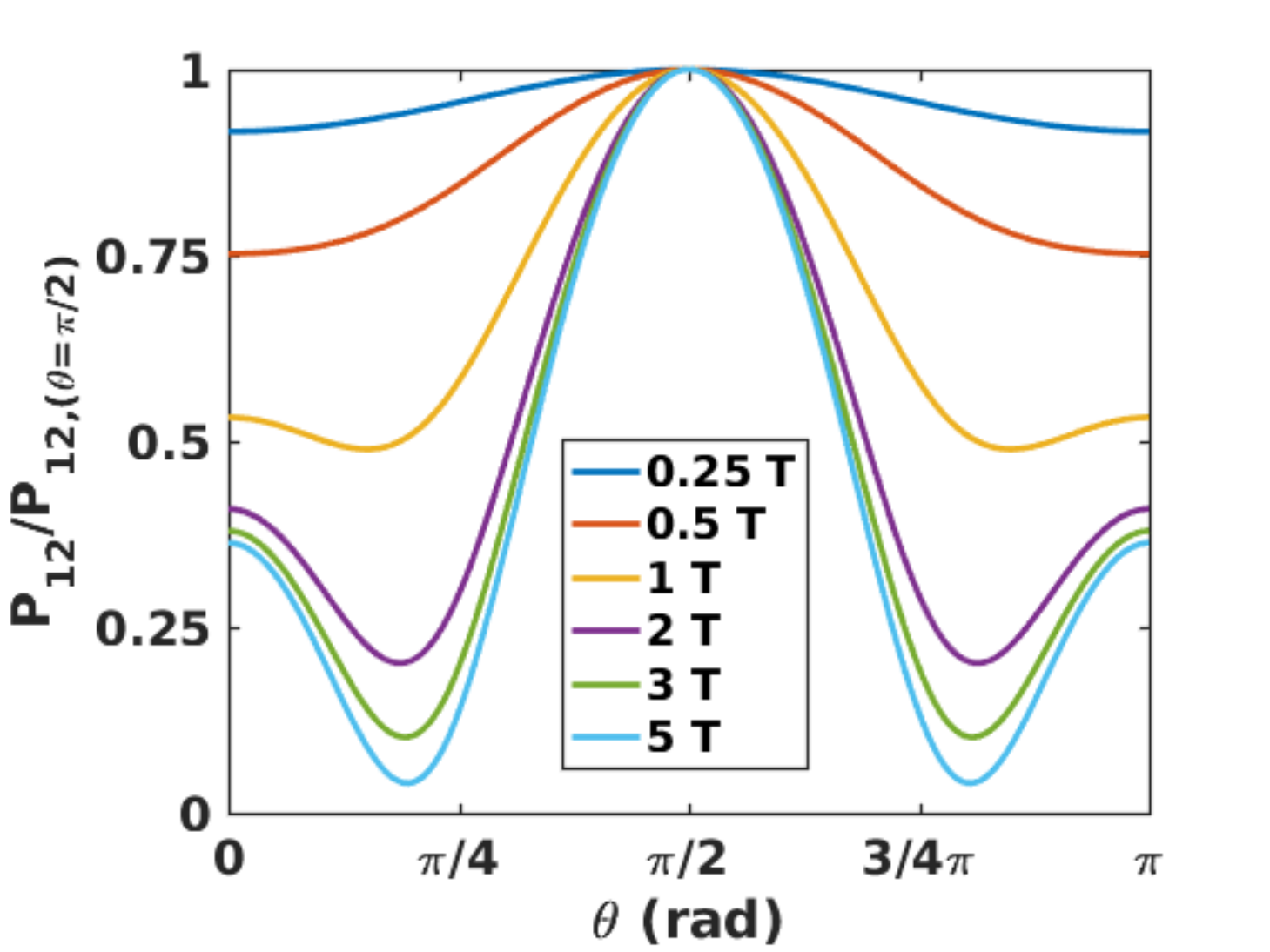}
	\caption{Interparticle heat flux $\langle P_{12} \rangle$ as a function of the angle between the magnetic field and the $z$-axis normalized to $\langle P_{12} \rangle (\theta = \pi/2)$, i.e. to the heat flux for a magnetic field parallel to the line connecting the two particles. Here we chose the same parameters as in Fig.~\ref{SpectralGiantResist}(a). }
	\label{MagneticFieldDirection}
\end{figure}

Figure~\ref{MagneticFieldDirection} shows the dependence of the exchanged flux on the angle $\theta$, for different magnitudes of the magnetic field, highlighting mainly two effects. First, when the magnitude of the magnetic field is increased the magnitude of the interparticle heat flux drops for all angles. This is the previously discussed giant magneto-resistance effect. Second, the heat flux has a global maximum for $\theta = \pi/2$ and a second local maximum for $\theta = 0$. In between the heat flux goes through the global minimum which is close to $\theta = \pi/4$ for strong magnetic fields. The position and existence of the minimum depend on the magnitude of the applied field, and on the interparticle distance. It is remarkable that the heat flux can be reduced by more than 90\% by changing the direction of the magnetic field. This effect is particularly pronounced close to $\theta = \pi/4$. On the contrary, when the direction of the magnetic field is parallel to the axis conecting the particles then no magneto-resistance effect exists. 

\section{Persistent heat current}\label{Sec_Persistent}

Now, we turn to the first configuration in which the non-reciprocity affects the heat current by producing an asymmetry in the heat flux. Starting from the general expression~\eqref{waermenetto}, when the material system is in local equilibrium at a given temperature $T_j$, it follows that the particles $i$ and $j$ exchange an energy flux
\begin{equation}
\langle P_{ij}^{\text{(eq)}}\rangle = 3\int_{0}^{\infty}\frac{\rd\omega}{2\pi}\,\Theta(\omega,T_j)[\mathcal{T}_{ij}(\omega,\mathbf{B})-\mathcal{T}_{ji}(\omega,\mathbf{B})]\label{Eq:equilibriumHeatFlux}.
\end{equation}
Hence, this exchange of energy is a direct signature of non-reciprocity, in the sense that it is different from zero if and only if $\mathcal{T}_{ij}\neq\mathcal{T}_{ji}$. Notice that since this the total net heat flux on each particle $i$ must vanish (i.e. $\sum_{j \neq i}^{N}\langle P_{ij}^{\text{(eq)}}\rangle =0$), the following general relation between the transmission coefficients holds~\cite{Latella2017}
\begin{equation}
 \underset{j}{\sum}[\mathcal{T}_{ij}(\omega,\mathbf{B})-\mathcal{T}_{ji}(\omega,\mathbf{B})]=0.\label{Eq:relation_coeffs}
\end{equation}
 It immediately follows from this relation that, despite the non-reciprocal behavior of the permittivity, a two-body  never displays an asymmetric heat flux, since we necessarily have $\mathcal{T}_{12}=\mathcal{T}_{21}$. Consequently, a configuration showing an asymmetrical flux must consist of at least three particles. Therefore, let us consider three particles sitting at the corners of an equilateral triangle. The applied magnetic field is taken to be perpendicular to plane common to the three particles. In this configuration non-reciprocity results indeed in a flux asymmetry, since $\mathcal{T}_{12} = \mathcal{T}_{23} = \mathcal{T}_{31} \neq \mathcal{T}_{13} = \mathcal{T}_{32} = \mathcal{T}_{21}$. In other words, the heat currents in clockwise and counterclockwise directions are different when applying a magnetic field. As a consequence, even for a local equilibrium situation where $T_1 = T_2 = T_3 = 300\,{\rm K}$ there is a persistent directional current as first pointed out by Zhu and Fan in Ref.~\cite{Zhu2016}.

\begin{figure}
	\centering
	\includegraphics[width=0.45\textwidth]{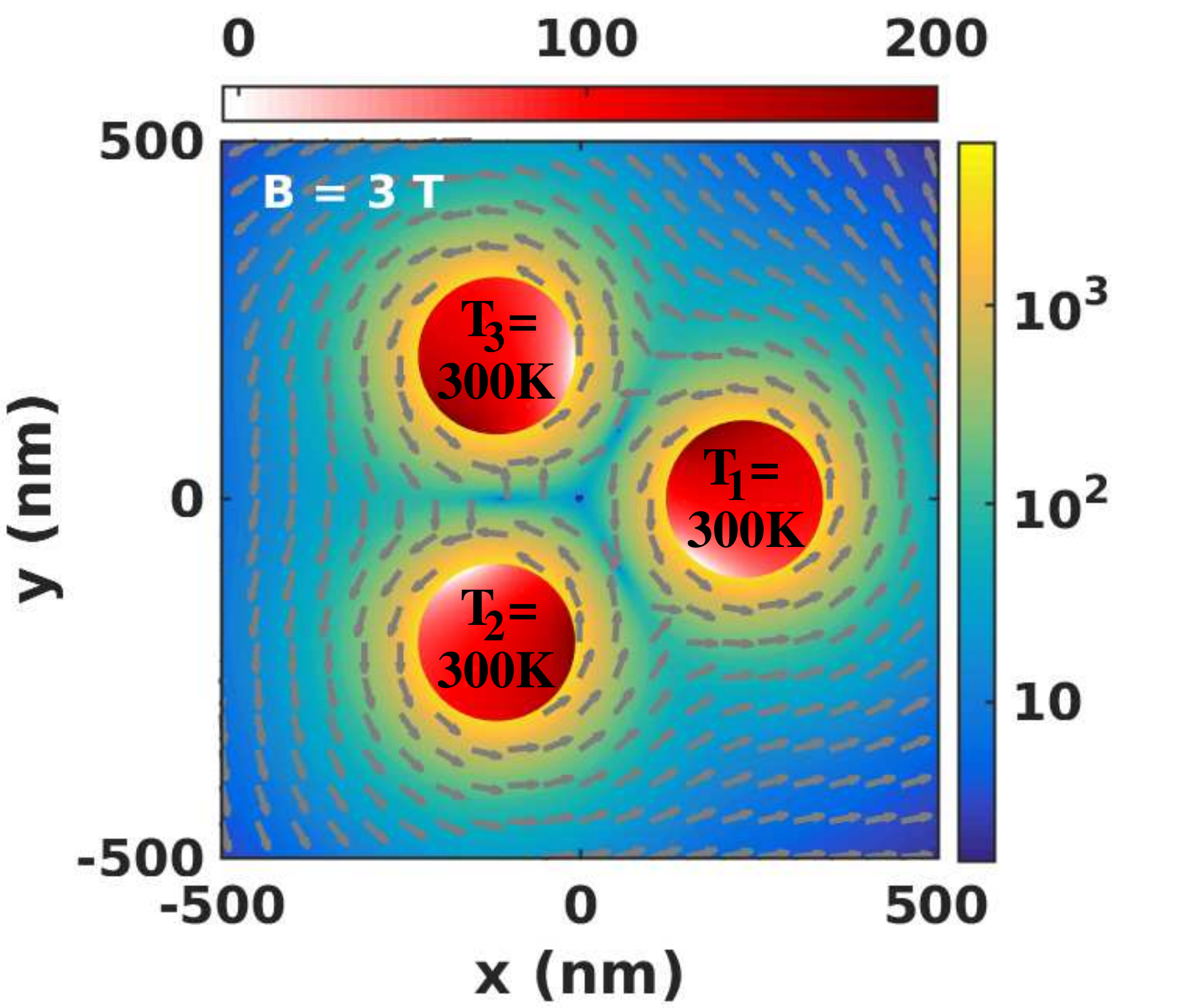}
	\caption{Persistent heat current in a three-particle configuration with $C_3$ symmetry, an interparticle distance of $400\,{\rm nm}$ and a magnetic field in $z$ direction with magnitude $B = 3\,$T. The vectors are the normalized Poynting vectors around the particles. The colorbar on the right-hand side gives the magnitude of the Poynting vector (W/m$^2$), while the colorbar on top gives the magnitude of the normal component of the Poynting vector (W/m$^2$) on the surface of the nanoparticles.} 
	\label{Persistentcurrent}
\end{figure}

 In Fig.~\ref{Persistentcurrent} we show the Poynting vector when $T_1 = T_2 = T_3 = 300\,{\rm K}$ in the presence of a magnetic field along the positive $z$ direction. Since the environment is at zero temperature, the particles mainly emit heat toward this environment. Due to the presence of the magnetic field, we have a circular heat flux in counterclockwise direction around the nanoparticles, resulting from the fact that the resonance with $m = +1$ dominates the heat flux in this configuration. Nonetheless, when calculating the interparticle heat current we find that 
\begin{equation}
  \langle P_{12} \rangle = \langle P_{23} \rangle = \langle P_{31} \rangle <  \langle P_{13} \rangle = \langle P_{32} \rangle = \langle P_{21} \rangle.
\end{equation}
Hence here we find a dominant heat current in the clockwise direction. We stress again that this does not mean that there is a net flux emitted or received by the particles, since it is easy to show that in local equilibrium we have
\begin{equation}
  \langle P_{1} \rangle = \langle P_{2} \rangle = \langle P_{3} \rangle = 0.
\end{equation}
That means that the particles stay at their temperatures $T_1 = T_2 = T_3$. Therefore, this persistent current does not result in a measurable heating or cooling. 

\begin{figure}
	\centering
	\includegraphics[width=0.45\textwidth]{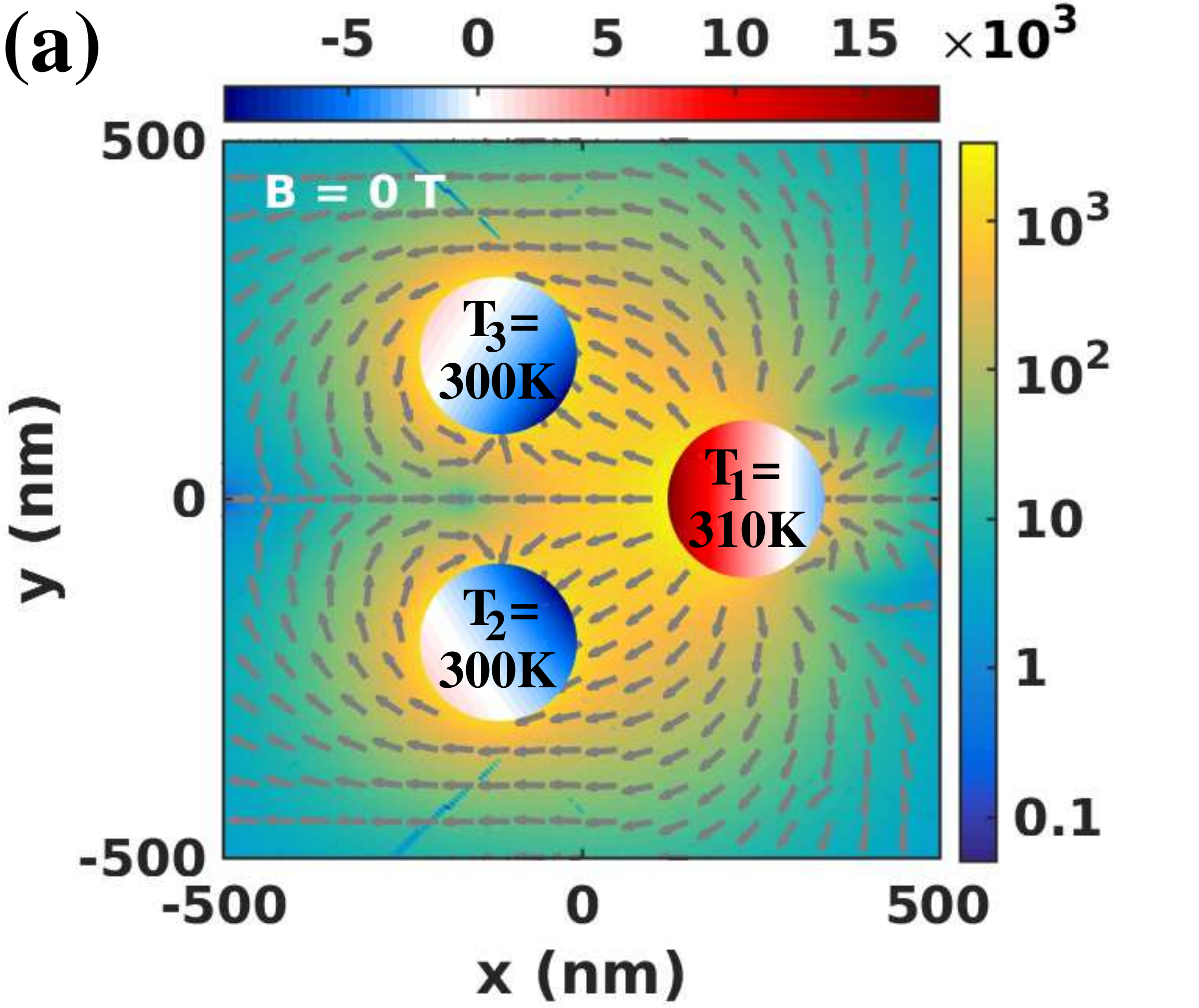}
	\includegraphics[width=0.45\textwidth]{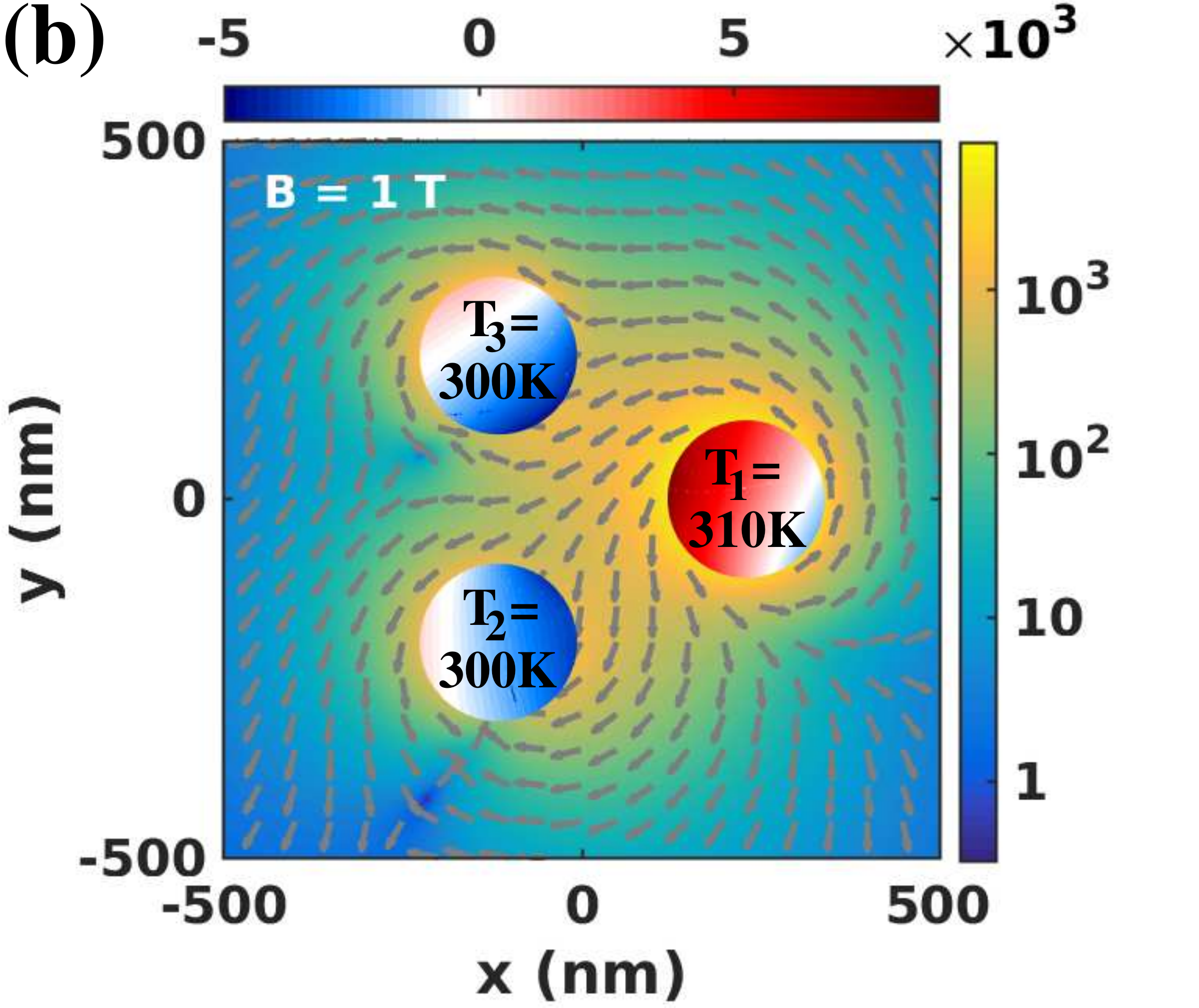}
        \includegraphics[width=0.45\textwidth]{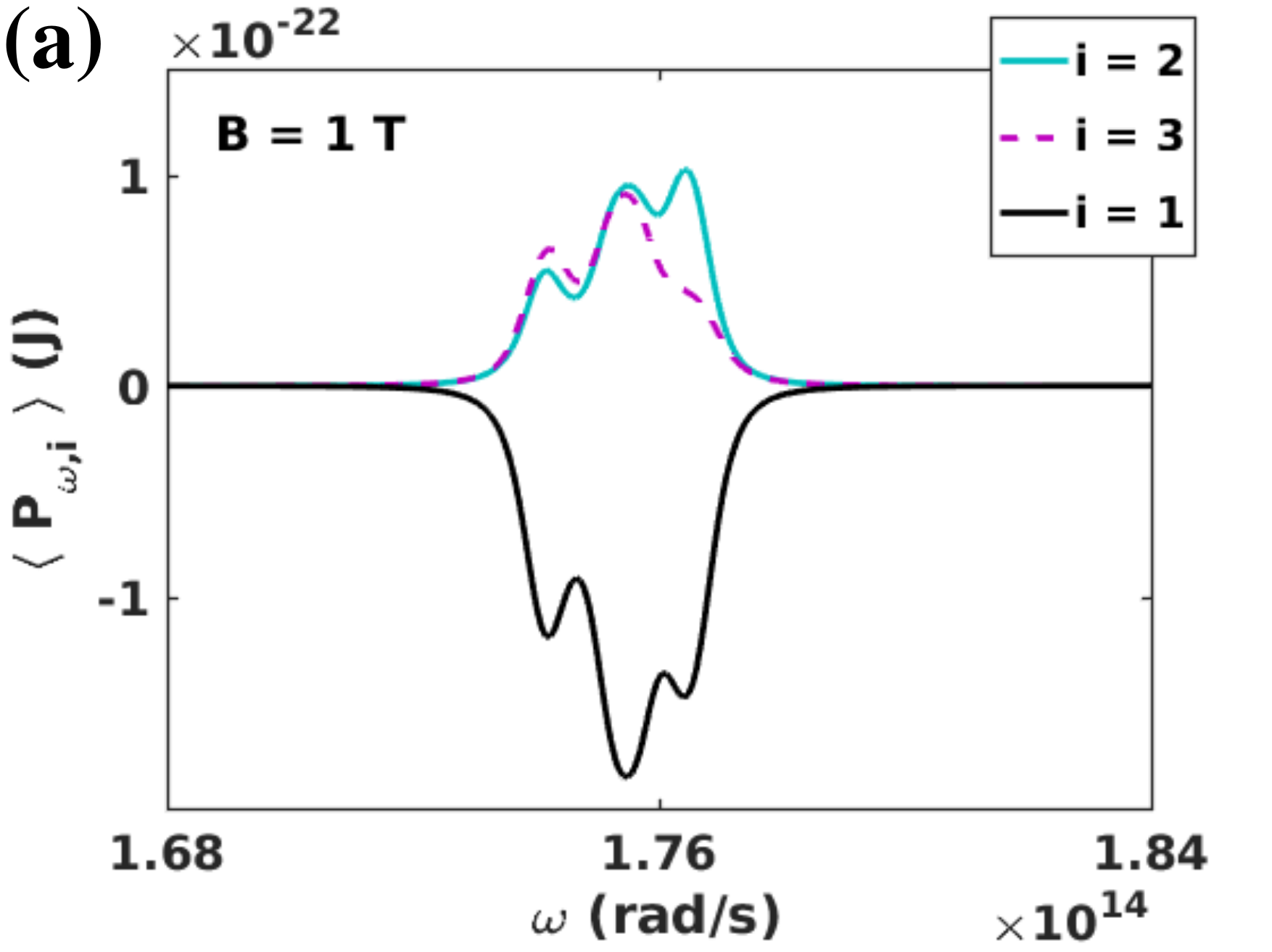}
	\caption{(a)-(b) Heat flux in a three particle configuration with one particle heated up with respect to the others for a magnetic field in $z$ direction with magnitude $B = 0\,$T and $1\,$T. (c) Spectral power  $ P_{\omega,i}$ ($i = 1,2,3$) emitted or received by each particle.} 
	\label{Persistentcurrent2}
\end{figure}

It is not obvious that the heat current is preferentially in clockwise direction, because the analysis of Poynting flux lines in Fig.~\ref{Persistentcurrent} seems to suggest a heat current in counterclockwise direction. To make the path which the heat flow takes more obvious we show in Fig.~\ref{Persistentcurrent2} the results of the heat flux when heating up particle $1$ to $T_1 = 310\,{\rm K}$ while keeping the other particles at $300\,{\rm K}$. It can be seen that, when applying the magnetic field, due to the circularity of the heat flux the path of heat flow bends toward particle $2$ so that the heat goes preferentially from particle $1$ to particle $2$. This is the reason for the heat current in clockwise direction which also persists when the temperatures of the particles are the same. To substantiate this we show in Fig.~\ref{Persistentcurrent2} also the spectral power $ P_{\omega,i}$ ($i = 1,2,3$). Then one can ses that particle 2 is more efficiently heated up than particle 1, indicating that the heat flux goes preferentially in the counterclockwise direction because of the dominant contribution of the $m = -1$ resonance.

\section{Thermal Hall effect}\label{Sec_Hall}

\begin{figure}
	\centering
	 \includegraphics[width=0.45\textwidth]{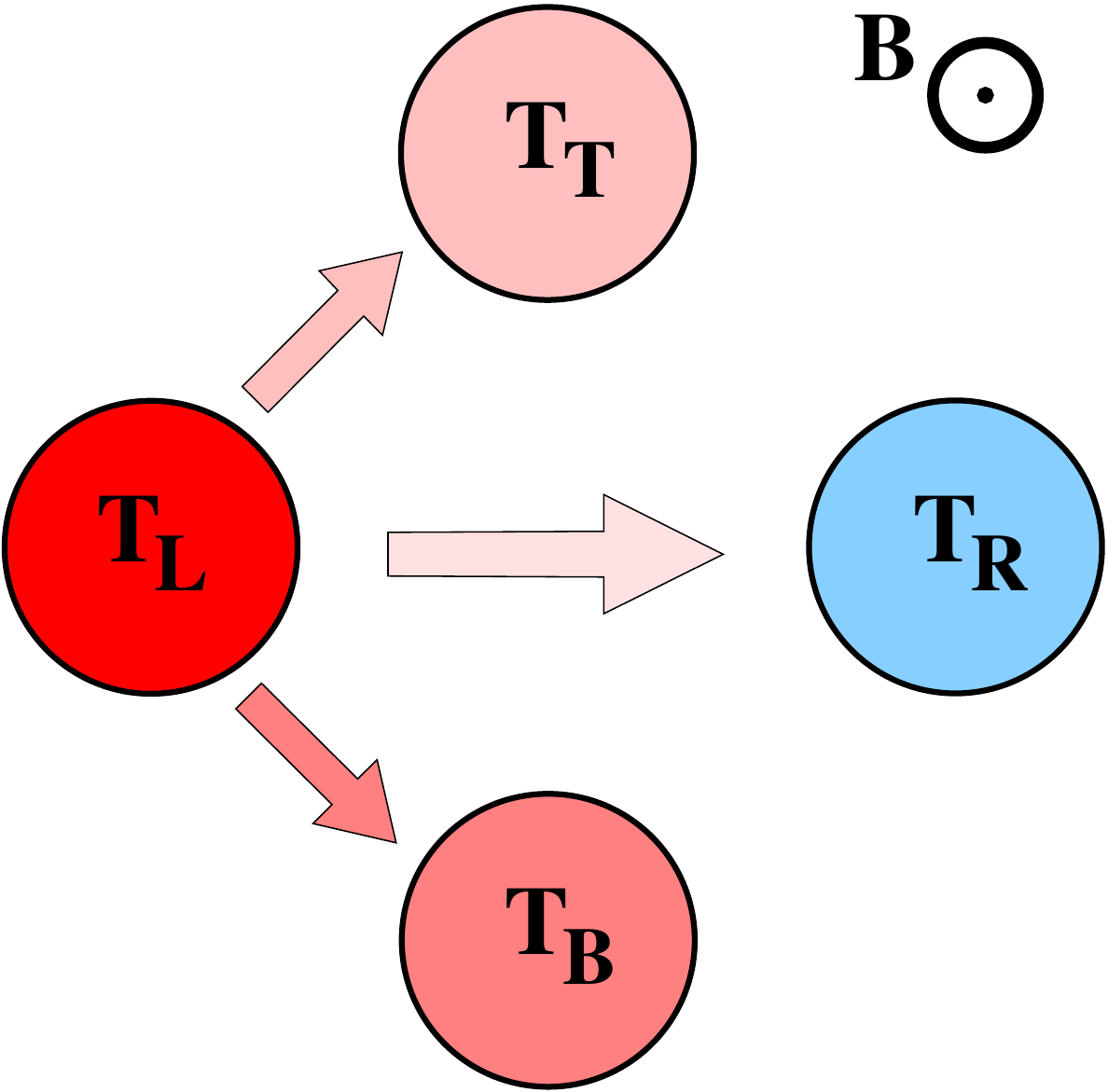}
	\caption{Sketch of the four-particle configuration to study the thermal Hall effect. The configuration has a discrete $C_4$ symmetry. The magnetic field is applied in the $z$ direction, i.e. perpendicular to the $x$-$y$ plane in which the particles are located.}
	\label{MOparticleHall}
\end{figure}

Let us now turn to the thermal Hall or Righi-Leduc effect for heat radiation. This effect has been first studied by Ben-Abdallah~\cite{Ben-Abdallah2016} within a four-particle configuration as depicted in Fig.~\ref{MOparticleHall}. In this configuration the transmission coefficients fulfill the properties
\begin{equation}
 \mathcal{T}_{LB} = \mathcal{T}_{BR} = \mathcal{T}_{RT} = \mathcal{T}_{TL} \equiv \mathcal{T}_{\rm clw},
\end{equation}
and 
\begin{equation}
 \mathcal{T}_{LT} = \mathcal{T}_{TR} = \mathcal{T}_{RB} = \mathcal{T}_{BL} \equiv \mathcal{T}_{\rm cclw}.
\end{equation}
Hence we have in principle one transmission coefficient describing the heat transfer between next neighbors in clockwise direction $\mathcal{T}_{\rm clw}$ and 
one transmission coefficient describing the heat transfer between next neighbors in counterclockwise direction $\mathcal{T}_{\rm cclw}$. As a consequence of non-reciprocity, we have in general $\mathcal{T}_{\rm clw} \neq \mathcal{T}_{\rm cclw}$. Furthermore, due to the symmetry of the configuration we further have 
\begin{equation}
 \mathcal{T}_{LR} = \mathcal{T}_{RL} = \mathcal{T}_{BT} = \mathcal{T}_{TB}, 
\end{equation}
i.e. the transmission coefficients describing the heat transfer between opposite particles are the same.

In order to study the thermal Hall effect it is assumed that the particle on the left and the one on the right are thermalized to $T_L = 310\,{\rm K}$ and $T_R = 300\,{\rm K}$, respectively. Then one lets the particles on the top and bottom dynamically go to their steady state. It is clear that without magnetic field the temperatures $T_B$ and $T_T$ of the particles at the top and bottom will reach a value which is between $T_L$ and $T_R$. Because of the symmetry of this configuration it is also clear that the steady-state temperatures will be equal, i.e. $T_B = T_T$. Now, when turning on the external magnetic field $\mathbf{B} = B \mathbf{e}_z$ it was shown that either the particle on the top or the one at the bottom becomes slightly hotter than the other one. Hence, in steady state a finite temperature difference $|T_B - T_T|$ perpendicular to the applied temperature difference $T_L - T_R$ is established. This is the thermal Hall or Righi-Leduc effect which has been explained in terms of the rotation of the optical axes of the nanoparticles due to the presence of the magnetic field. Based on the previous discussion, it is evident that we can now understand it as a consequence of the circular heat flux which is itself an effect of the Lorentz force acting on the electrons inside each particle.

\begin{figure}
	\centering
        \includegraphics[width=0.45\textwidth]{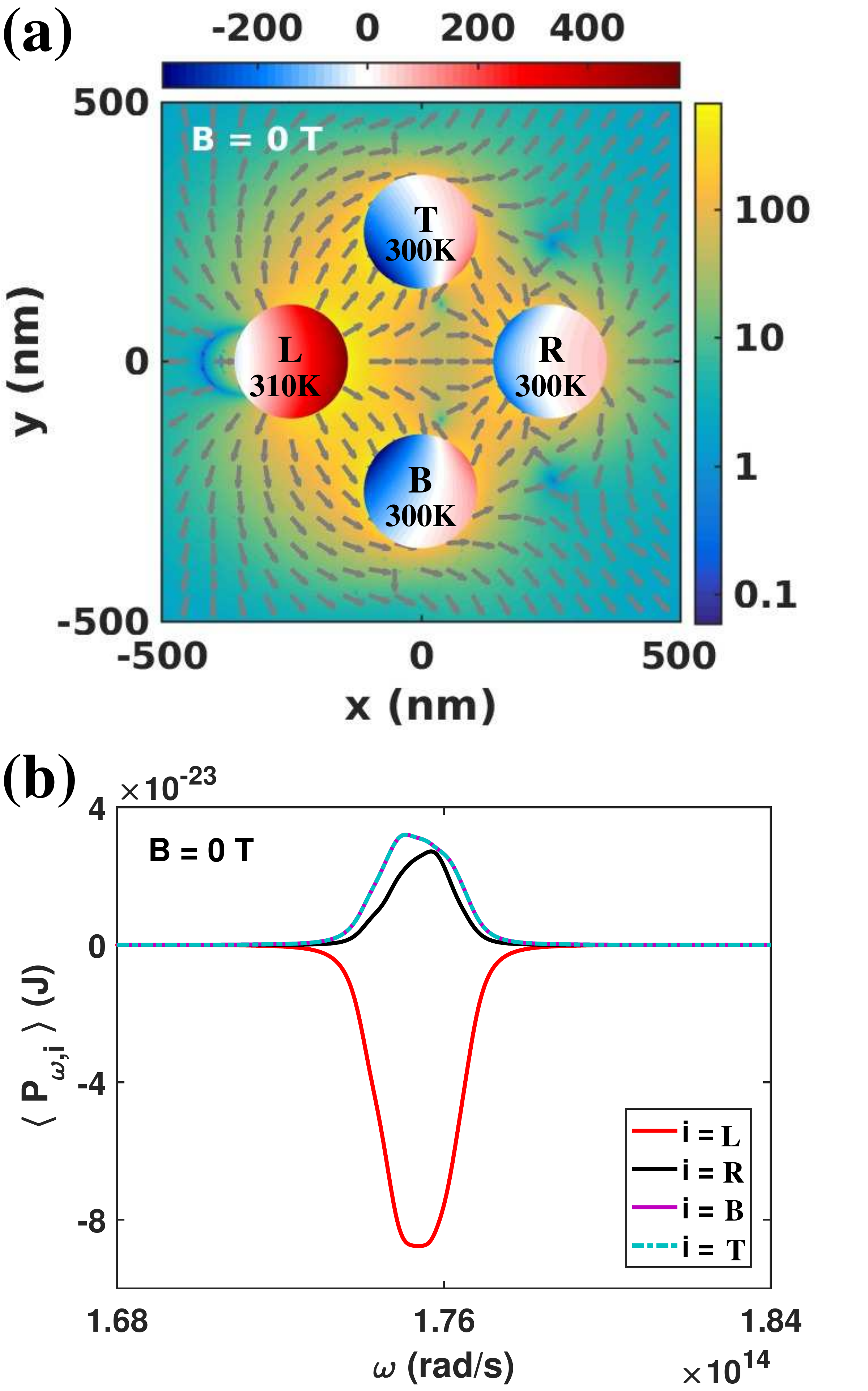}
	\caption{(a) Heat flux in a four-particle configuration with an interparticle distance 
         of $500\,{\rm nm}$ with fixed temperatures and no applied magnetic field. 
         The vectors are the normalized Poynting vectors around the particles. The colorbar on the right-hand side gives the magnitude of the Poynting vectors (W/m$^2$), while the colorbar on top gives the magnitude of the normal component of the Poynting vector (W/m$^2$) on the surface of the nanoparticles.
         (b) Spectral power $ P_{\omega, L/R/T/B}$ emitted or received by each particle.} 
	\label{Hall0T}
\end{figure}

For a quantitative study of the thermal Hall effect we first consider the case where the particle on the left has a temperature of $T_L = 310\,{\rm K}$ and the other particles have the fixed temperature of $300\,{\rm K}$. In Fig.~\ref{Hall0T} we first show the heat flux around the particles and the power emitted or received by them when no magnetic field is applied. It can be nicely seen in Fig.~\ref{Hall0T}(a) that the Poynting vector field is symmetric with respect to the $x$-axis. Therefore the heat flow to the upper and lower particle is the same. This can also be verified by the fact that the spectral power received by the particle at the bottom and by the one at the top are exactly the same for all frequencies [see Fig.~\ref{Hall0T}(b)]. 

\begin{figure*}
	\centering
    \includegraphics[width=0.45\textwidth]{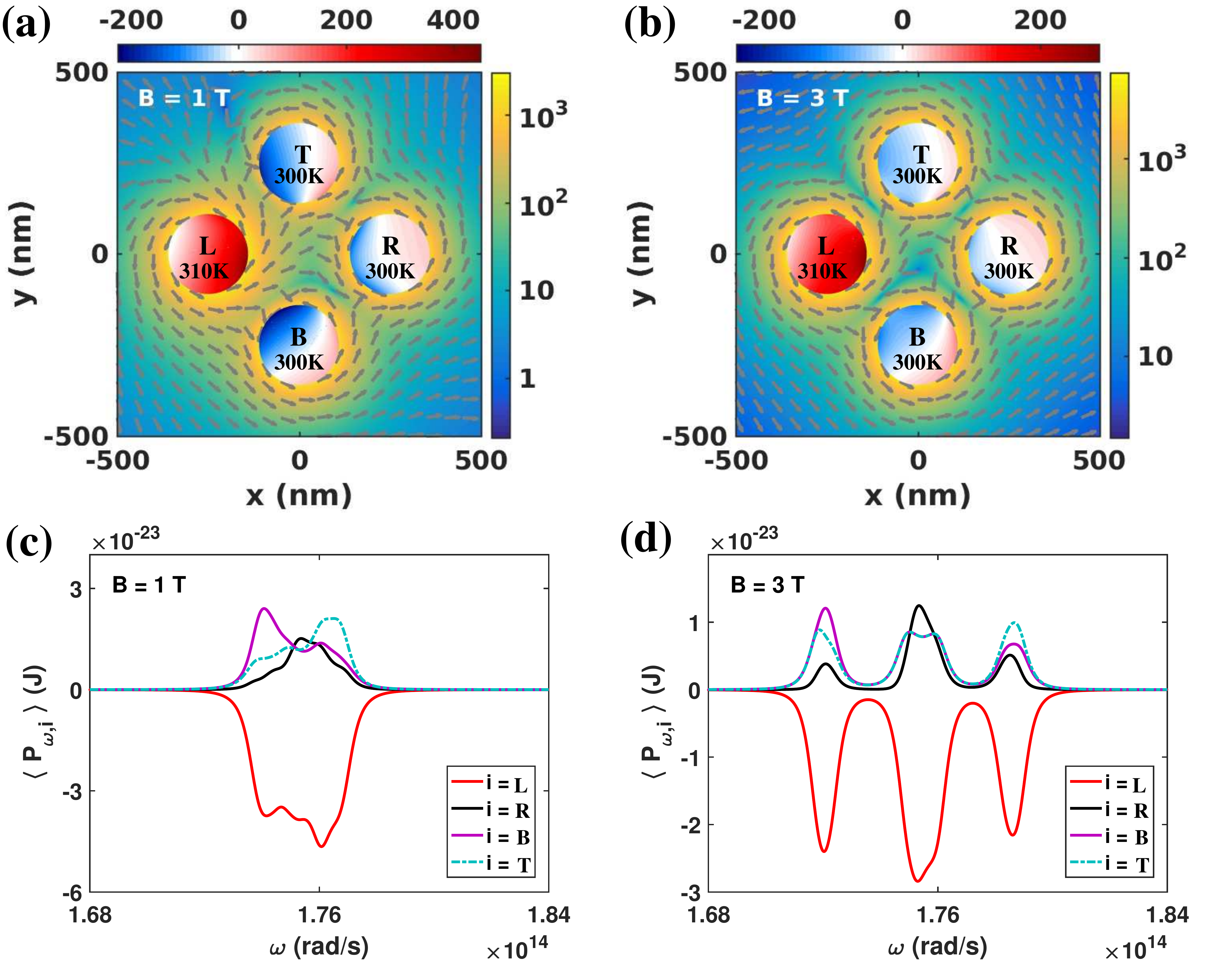}
	\caption{(a)-(b) Heat flux in a four-particle configuration with an interparticle distance 
         of $500\,{\rm nm}$, fixed temperatures and an applied magnetic field in $z$ direction of $B = 1\,{\rm T}$ and $B = 3\,{\rm T}$. The vectors are the normalized Poynting vectors around the particles. The colorbar on the right-hand side gives the magnitude of the Poynting vectors (W/m$^2$), while the colorbar on top gives the magnitude of the normal component of the Poynting vector (W/m$^2$) on the surface of the nanoparticles. (c)-(d) Spectral power $\langle P_{L/R/T/B, \omega} \rangle$ received by each particle for $B = 1\,{\rm T}$ and $B = 3\,{\rm T}$.}
	\label{Hall1T3T}
\end{figure*}

When applying the magnetic field in $z$ direction the heat flux becomes circular as shown in Fig.~\ref{Hall1T3T}(a). It is tempting to assume that, due to the circularity of the heat flux in counterclockwise direction (associated to a more pronounced contribution of the resonance for $m = +1$), the heat current is counterclockwise as well. We have seen before, that this is not necessarily the case, because the counterclockwise circularity adds up into a clockwise circular heat current in the inner part of the four-particle configuration. Furthermore, there is a competition between the heat currents associated to the $m = +1$ and $m = -1$ resonances. This can be seen in the spectral power received by the upper and lower particle in Fig.~\ref{Hall1T3T}(c). For $B = 1\,{\rm T}$ the lower particle is mainly heated up by the $m = +1$ resonance and the upper particle is heated up by the $m = -1$ resonance. It seems that the $m = +1$ resonance in $\langle P_{B,\omega} \rangle$ is slightly stronger than the $m = -1$ resonance in $\langle P_{T,\omega} \rangle$. This suggests that the particle at the bottom is heated up more efficiently than the particle at the top. When integrating the spectral power, it turns out that this is indeed the case.

\begin{figure}
	\centering
	\includegraphics[width=0.45\textwidth]{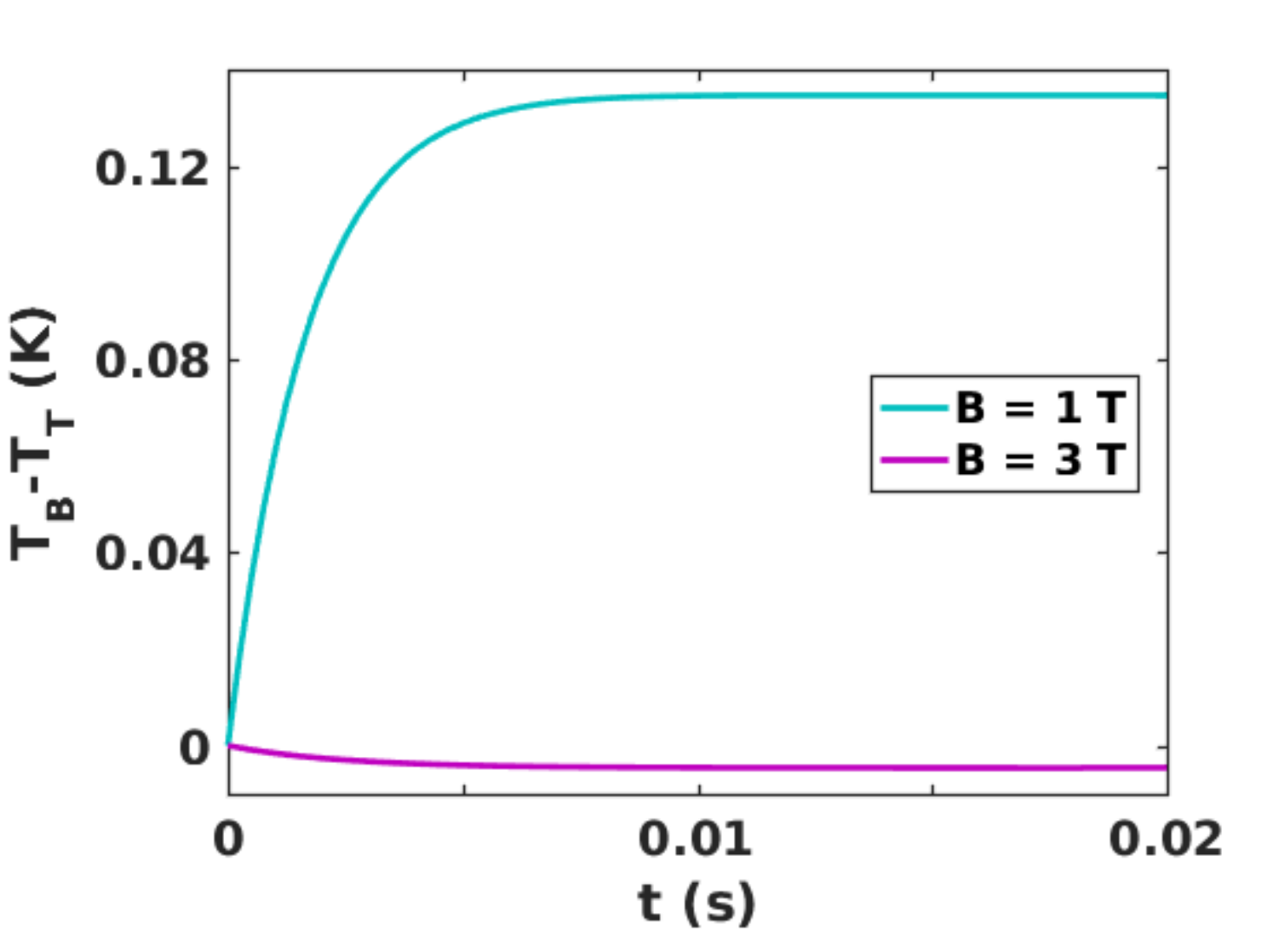}
	\caption{Dynamical evolution of the temperatur difference $T_B(t) - T_T(t)$ for the four-particle configuration with particles having a radius of $100\,{\rm nm}$ and an interparticle distance of $500\,{\rm nm}$. At the initial time the temperatures are as in Figs.~\ref{Hall0T} and \ref{Hall1T3T}. Furthermore the temperatures of the left and right particles $T_L$ and $T_R$ are fixed.}
	\label{RelDyn}
\end{figure}

Nonetheless, in Fig.~\ref{Hall1T3T}(c) there is for $B = 1\,{\rm T}$ also a power flow to the lower particle by the $m = -1$ resonance and to the upper particle by the $m = +1$ resonance. This corresponds to a clockwise or counterclockwise rotation of the heat current within the four-particle system. When increasing the magnitude of the magnetic field to the value $3\,{\rm T}$, the heating of the lower particle by the $m = -1$ resonance and the heating of the upper particle by the $m = +1$ resonance become stronger. In this case it is not directly evident from the spectral power plotted in Fig.~\ref{Hall1T3T}(d) which particle will receive more heat. When integrating the spectral power we find that for $B = 3\,{\rm T}$ the particle on top heats up more efficiently than the particle at the bottom. 

\begin{figure}
	\centering
	\includegraphics[width=0.45\textwidth]{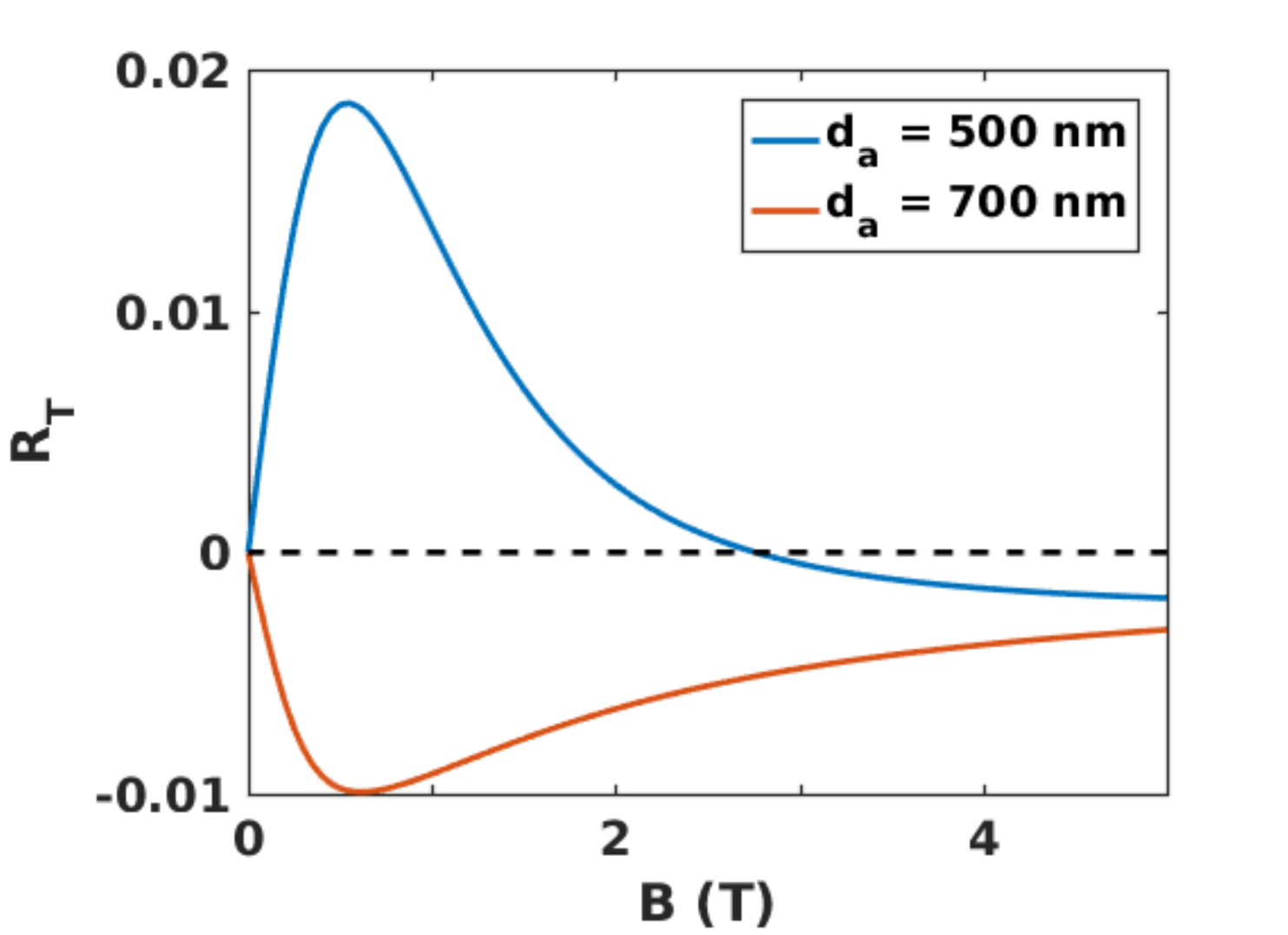}
	\caption{Relative Hall temperature difference defined in Eq.~\eqref{Eq:RLC} as a function of the magnetic field strength for different interparticle distances $d_a = 500\,{\rm nm}$ and $700\,{\rm nm}$.}
	\label{RuberB}
\end{figure}

In order to study the directionality of the heat flow in the four-particle configuration or the thermal Hall effect in more detail, we now assume that at time $t = 0$ the particles have the same temperatures as above, i.e. $T_L = 310\,{\rm K}$ and the other particles have a temperature of $300\,{\rm K}$. The difference is now that we fix the temperatures of the left and the right particles at $T_L = 310\,{\rm K}$ and $T_R = 300\,{\rm K}$, and let the temperatures of the upper and lower particle evolve in time until they reach their steady-state value. This dynamic process can be described by the set of two differential equations $(k = B,T)$
\begin{equation}
  \rho C V \frac{dT_k}{dt} = \langle P_{k}\bigl(T_L,T_R,T_B(t), T_T(t)\bigr)\rangle, 
\label{dgltemp}
\end{equation}
where the values for the heat capacity is $C = 200\,$J/kg K and the mass density $\rho = 5775\,$kg/m$^3$ of InSb are taken from Ref.~\cite{Piesbergen1963}, and $V$ is the volume of the nanoparticles. This system of differential equations can be easily solved by a Runge-Kutta method. In Fig.~\ref{RelDyn} we plot the temporal evolution of the temperature difference $T_B - T_T$ for two magnetic field strengths. It can be seen that the steady state is reached in our configuration after ten milliseconds. Furthermore, it can be seen that for $B = 1\,{\rm T}$ the particle at the bottom is hotter than the particle at the top in the steady state. For $B = 3\,{\rm T}$ we find the opposite result. Hence the direction of the thermal Hall effect depends on the magnetic field strength.

This dependence of the steady-state temperatures on the magnetic field strength and therefore the directionality of the thermal Hall effect can be better studied by introducing the relative Hall temperature difference defined by
\begin{equation}
  R_T = \frac{T_B^{\text{(st)}} - T_T^{\text{(st)}}}{T_L - T_R},
\label{Eq:RLC}
\end{equation} 
where $T_B^{\text{(st)}}$ and $T_T^{\text{(st)}}$ are the steady-state temperatures of the bottom and top particle, respectively, defined by the condition $\rd T_{B/T} /\rd t = 0$. In Fig.~\ref{RuberB} we show $R_T$ as a function of the mangnetic field strength for two different interparticle distances. This plot confirms that for $d_a = 500\,{\rm nm}$ and weak magnetic fields $R_T > 0$ so that the particle at the bottom is heated up more efficiently than the particle at the top. Hence for small magnetic fields the directionality of the Hall effect follows the directionality of the circular heat flux around the particles which is in counterclockwise direction. Interestingly, for magnetic fields stronger than $2.7\,{\rm T}$ the directionality changes and the particle at the top is heated up more efficiently. It is interesting to remark that that the directionality of the heat flow also depends on the the geometrical configuration. As a matter of fact, when chosing $d_a = 700\,{\rm nm}$ we find already the reverse effect. This reversal has to be expected from our discussion in Sec.~\ref{Sec_Circular} to happen at the transition from near field to far field, where the $m = -1$ resonance dominates. Here this transition from a $m = +1$ to a $m = -1$ dominated heat flux happens already for a relatively small distance which is due to the different parameter-set A used for the Hall effect. If we used the parameter-set B instead, this transition would happen at larger distances. Furthermore, we find that for the parameter-set B the Hall effect is rather small. 

\section{Thermal gradient and magnetic field sensing}\label{Sec_Sensing}

Before concluding, let us briefly discuss some potential applications of thermal Hall effect in the field of temperature or magnetic-field sensing. Let us start by considering a system analog to the four-terminal junction made with magneto-optical nanoparticles used to demonstrate the existence of a photon Hall effect (Fig.~\ref{MOparticleHall}). Under the action of an external magnetic field of weak intensity and a relatively weak temperature gradient $\Delta T$ between the left and right particle, the Hall flux exchanged between the bottom and top particles reads
\begin{equation}
 \varphi_H= G_H \Delta_H T=G_H R_T \Delta T=\beta B,
\label{Eq:sensor1}
\end{equation} 
where $ \Delta_H T=T_B - T_T$, $ \Delta T=T_L - T_R$, $G_H$ is the Hall thermal conductance which can be easily calculated knowing the geometric configuration and the optical properties of particles, and $\beta$ is a proportionality coefficient. Therfore on one hand a simple meaure of $ \Delta_H T$ gives us the Hall flux $ \varphi_H$ and also the magnetic-field intensity. On the other hand, if the applied magnetic field and the temperature gradient $\Delta T$ are known then relation~\eqref{Eq:sensor1} gives the value of transversal gradient $ \Delta_H T$. 

\section{Summary}\label{Sec_Summary}

We have reviewed some recently described basic mechanisms which drive the radiative heat exchanges in systems made with many magneto-optical particles. In the particular case of a single particle, we have shown that one can observe a persistent heat flux 
 which is very similar to persistent heat current described in more complex situations and which is fundamentally related to the singular topological structure of heat flux. Based on the discussion of the circular heat flux around a magneto-optical particle we have tried to intuitively understand the directionality of the heat flux in many-particle configurations
with three and four particles. We have shown that this directionality can in principle be understood by the circular heat flux but it highly depends on the configuration and the magnetic-field strength. In particular, for three particles in an equilateral triangle the persistent heat current and also the heat current in general seems to be in the opposite direction of the circularity of the heat flux around the particles. However, in 
a symmetric four-particle configuration as the one considered for the thermal Hall effect, for small magnetic fields the directionality is in the same direction as the circular heat flux around the particles. We have seen that due to the competing contributions of the $m = +1$ and $m = -1$  resonances it is not always {\itshape a priori} clear in which direction the heat will flow.  Besides, we have shown that the non-reciprocity can be exploited to make thermal and magnetic field sensing and to modulate the heat flow in multiple directions. 

\newpage

\acknowledgments
\noindent

The authors acknowledge financial support by the DAAD and Partenariat Hubert Curien Procope Program (project 57388963). S.-A.\ B. acknowledges support from Heisenberg Programme of the Deutsche Forschungsgemeinschaft (DFG, German Research Foundation) under the project No. 404073166.


\newpage



\begin{thebibliography}{999}
  \bibitem{YangWang2017} Y. Yang and L. Wang, "Electrically-controlled near-field radiative thermal modulator made of graphene-coated silicon carbide plates," {\itshape J. Quant. Spec. Rad. Transf.}, {\bf 197}, pp. 68-75 (2017) [doi: 10.1016/j.jqsrt.2016.06.013].
  \bibitem{HuangEtAl2014} Y. Huang, S. V. Boriskina, and G. Chen, "Electrically tunable near-field radiative heat transfer via ferroelectric materials," {\itshape Appl. Phys. Lett.} {\bf 105}, 244102 (2014) [doi: 10.1063/1.4904456].
  \bibitem{PBASAB2015} P. Ben-Abdallah and S.-A. Biehs, "Contactless heat flux control with photonic devices," {\itshape AIP Advances} {\bf 5}, 053502 (2015) [doi: 10.1063/1.4915138].
  \bibitem{PBASAB2013} P. Ben-Abdallah and S.-A. Biehs, "Phase-change radiative thermal diode," {\itshape Appl. Phys. Lett.} {\bf 103}, 191907 (2013) [doi: 10.1063/1.4829618].
  \bibitem{vanZwol1} P. van Zwol, K. Joulain, P. Ben-Abdallah, and J. Chevrier, "Phonon polaritons enhance near-field thermal transfer across the phase transition of VO$_2$," {\itshape Phys. Rev. B} \textbf{84}, 161413(R) (2011) [doi: 10.1103/PhysRevB.84.161413].
  \bibitem{NefzaouiEtAl2014} E. Nefzaoui, J. Drevillon, Y. Ezzahri, and K. Joulain, "Radiative thermal rectification using superconducting materials," {\itshape Appl. Phys. Lett.} {\bf 104}, 103905 (2014) [doi: 10.1063/1.4868251].
  \bibitem{Kubytskyi2014} V. Kubytskyi, S.-A. Biehs, and P. Ben-Abdallah, "Radiative Bistability and Thermal Memory," {\itshape Phys. Rev. Lett.} {\bf 113}, 074301 (2014) [doi: 10.1103/PhysRevLett.113.074301].
  \bibitem{DyakovEtAl2015} S. A. Dyakov, J. Dai, M. Yan, and M. Qiu, "Near field thermal memory based on radiative phase bistability of VO$_2$," {\itshape J. Phys. D: Appl. Phys.} {\bf 48} 305104 (2015) [doi: 10.1088/0022-3727/48/30/305104].
  \bibitem{PBASAB2014} P. Ben-Abdallah and S.-A. Biehs, "Near-Field Thermal Transistor," {\itshape  Phys. Rev. Lett.} {\bf 112}, 044301 (2014) [doi: 10.1103/PhysRevLett.112.044301].
  \bibitem{OrdonezEtAl2016} J. Ordonez-Miranda, Y. Ezzahri, J. Drevillon, and K. Joulain, "Transistorlike Device for Heating and Cooling Based on the Thermal Hysteresis of VO$_2$," {\itshape Phys. Rev. Appl.} {\bf 6}, 054003 (2016) [doi: 10.1103/PhysRevApplied.6.054003].
   \bibitem{PBASAB2016} P. Ben-Abdallah and S.-A. Biehs, "Towards Boolean operations with thermal photons," {\itshape Phys. Rev. B} {\bf 94}, 241401(R) (2016) [doi: 10.1103/PhysRevB.94.241401].
  
  \bibitem{ItoEtAl2014} K. Ito, K. Nishikawa, H. Iizuka, and H. Toshiyoshi, "Experimental investigation of radiative thermal rectifier using vanadium dioxide ," {\itshape Appl. Phys. Lett.} {\bf 105}, 253503 (2014) [doi: 10.1063/1.4905132].
  \bibitem{ItoEtAl2016} K. Ito, K. Nishikawa, and H. Iizuka, "Multilevel radiative thermal memory realized by the hysteretic metal-insulator transition of vanadium dioxide," {\itshape Appl. Phys. Lett.} {\bf 108}, 053507 (2016) [doi: 10.1063/1.4941405].
  \bibitem{FiorinoEtAl2018} A. Fiorino, D. Thompson, L. Zhu, R. Mittapally, S.-A. Biehs, O. Bezencenet, N. El-Bondry, S. Bansropun, P. Ben-Abdallah, E. Meyhofer, and P. Reddy, "A Thermal Diode Based on Nanoscale Thermal Radiation," {\itshape ACS Nano} {\bf 12}, pp 5774-5779 (2018) [doi: 10.1021/acsnano.8b01645].

\bibitem{OttEtAl2018} A. Ott, P. Ben-Abdallah, and S.-A. Biehs, "Circular heat and momentum flux radiated by magneto-optical nanoparticles," {\itshape Phys. Rev. B} {\bf 97}, 205414 (2018) [doi: 10.1103/PhysRevB.97.205414].
\bibitem{Zhu2016} L. Zhu and S. Fan, "Persistent Directional Current at Equilibrium in Nonreciprocal Many-Body Near Field Electromagnetic Heat Transfer," {\itshape Phys. Rev. Lett.} {\bf 117}, 134303 (2016) [doi: 10.1103/PhysRevLett.117.134303].
\bibitem{Latella2017} I. Latella and P. Ben-Abdallah, "Giant Thermal Magnetoresistance in Plasmonic Structures," {\itshape Phys. Rev. Lett.} {\bf 118}, 173902, (2017) [doi: 10.1103/PhysRevLett.118.173902].
\bibitem{Cuevas} R. M. Abraham Ekeroth, P. Ben-Abdallah, J.C. Cuevas, and A. Garcia Martin, "Anisotropic Thermal Magnetoresistance for an Active Control of Radiative Heat Transfer," {\itshape ACS Photonics} {\bf 5}, pp. 705-710 (2017) [doi: 10.1021/acsphotonics.7b01223]. 
\bibitem{Ben-Abdallah2016} P. Ben-Abdallah, "Photon Thermal Hall Effect," {\itshape Phys. Rev. Lett.} {\bf 116}, 084301, (2016) [doi: 10.1103/PhysRevLett.116.084301].

 \bibitem{PBAEtAl2011} P. Ben-Abdallah, S.-A. Biehs, and K. Joulain, "Many-Body Radiative Heat Transfer Theory," {\itshape Phys. Rev. Lett.} {\bf 107}, 114301 (2011) [doi: 10.1103/PhysRevLett.107.114301].
 \bibitem{MessinaEtAl2013} R. Messina, M. Tschikin, S.-A. Biehs, and P. Ben-Abdallah,"Fluctuation-electrodynamic theory and dynamics of heat transfer in systems of multiple dipoles," {\itshape Phys. Rev. B} {\bf 88}, 104307 (2013) [doi: 10.1103/PhysRevB.88.104307].
 \bibitem{DongEtAl2017} J. Dong, J. Zhao, and L. Liu, "Radiative heat transfer in many-body systems: Coupled electric and magnetic dipole approach," {\itshape Phys. Rev. B} {\bf 95}, 125411 (2017) [doi: 10.1103/PhysRevB.95.125411].
 \bibitem{EkerothEtAl2017} R. M. A. Ekeroth, A. Garcia-Martin, and J. C. Cuevas, "Thermal discrete dipole approximation for the description of thermal emission and radiative heat transfer of magneto-optical systems," {\itshape Phys. Rev. B} {\bf 95}, 235428 (2017) [doi: 10.1103/PhysRevB.95.235428].
 \bibitem{Saaskilathi2014} K. S\"{a}\"{a}skilathi, J. Oksanen, and J. Tulkki, "Quantum Langevin equation approach to electromagnetic energy transfer between dielectric bodies in an inhomogeneous environment," {\itshape Phys. Rev. B} {\bf 89}, 134301 (2014) [doi: 10.1103/PhysRevB.89.134301].
 \bibitem{Nikbakht2014} M. Nikbakht,"Radiative heat transfer in anisotropic many-body systems: Tuning and enhancement," {\itshape J. Appl. Phys.} {\bf 116},  094307 (2014) [doi: 10.1063/1.4894622].
 \bibitem{PBAEtAl2013} P. Ben-Abdallah, R. Messina, S.-A. Biehs, M. Tschikin, K. Joulain, and C. Henkel, "Heat Superdiffusion in Plasmonic Nanostructure Networks," {\itshape Phys. Rev. Lett.} {\bf 111}, 174301 (2013) [doi: 10.1103/PhysRevLett.111.174301].
  \bibitem{PHanEtAl2013} A. D. Phan, T.-L. Phan, and L. Woods, "Near-field heat transfer between gold nanoparticle arrays," {\itshape J. Appl. Phys.} {\bf 114}, 214306 (2013) [doi: 10.1063/1.4838875].
 \bibitem{Nikbakht2017} M. Nikbakht, "Radiative heat transfer in fractal structures," {\itshape Phys. Rev. B} {\bf 96}, 125436 (2017) [doi: 10.1103/PhysRevB.96.125436].
 \bibitem{Edalatpour2014} S. Edalatpour and M. Francoeur, "The Thermal Discrete Dipole Approximation (T-DDA) for near-field radiative heat transfer simulations in three-dimensional arbitrary geometries," {\itshape J. Quant. Spec. Rad. Trans.} {\bf 133}, pp.  364-373 (2014) [doi: 10.1016/j.jqsrt.2013.08.021].
\bibitem{Jose2018} J. Ordonez-Miranda, L. Tranchant, K. Joulain, Y. Ezzahri, J. Drevillon, and S. Volz, "Thermal energy transport in a surface phonon-polariton crystal," {\itshape Phys. Rev. B} {\bf 93}, 035428 (2016) [doi: 10.1103/PhysRevB.93.035428].
 \bibitem{PBA2008} P. Ben-Abdallah, K. Joulain, J. Drevillon, and C. Le Goff, "Heat transport through plasmonic interactions in closely spaced metallic nanoparticle chains," {\itshape Phys. Rev. B} {\bf 77}, 075417 (2008) [doi: 10.1103/PhysRevB.77.075417].
 \bibitem{Jose2015} J. Ordonez-Miranda, L. Tranchant, S. Gluchko, and S. Volz, "Energy transport of surface phonon polaritons propagating along a chain of spheroidal nanoparticles," {\itshape Phys. Rev. B} {\bf 92}, 115409 (2015) [doi: 10.1103/PhysRevB.92.115409].
 \bibitem{KathmannEtAl2018} C. Kathmann, R. Messina, P. Ben-Abdallah, and S.-A. Biehs, "Limitations of kinetic theory to describe near-field heat exchanges in many-body systems," {\itshape Phys. Rev. B} {\bf 98}, 115434 (2018) [doi: 10.1103/PhysRevB.98.115434].
  \bibitem{DongEtAl2018} J. Dong, J. Zhao, and L. Liu, "Long-distance near-field energy transport via propagating surface waves," {\itshape Phys. Rev. B} {\bf 97}, 075422 (2018) [doi: 10.1103/PhysRevB.97.075422].
 \bibitem{MessinaEtAl2018} R. Messina, S.-A. Biehs, and P. Ben-Abdallah, "Surface-mode-assisted amplification of radiative heat transfer between nanoparticles," {\itshape Phys. Rev. B} {\bf 97}, 165437 (2018) [doi: 10.1103/PhysRevB.97.165437].
 \bibitem{Callen} H. B. Callen and T. A. Welton, "Irreversibility and Generalized Noise," {\itshape Phys. Rev.} {\bf 83}, 34 (1951) [doi: 10.1103/PhysRev.83.34].
\bibitem{LakhtakiaEtAl1991} A. Lakhtakia, V. K. Varadan, and V. V. Varadan, "Low-frequency scattering by an imperfectly conducting sphere immersed in a dc magnetic field," {\itshape International Journal of Infrarared and Millimeter Waves} {\bf 12}, pp. 1253-1264 (1991) [doi: 10.1007/BF01014683].
\bibitem{Albaladejo} S. Albaladejo, R. G\'{o}mez-Medina, L. S. Froufe-P\'{e}rez, H. Marinchio, R. Carminati, J. F. Torrado, G. Armelles, A. Garc\'{i}a-Mart\'{i}n, and J. J. S\'{a}enz, "Radiative corrections to the polarizability tensor of an electrically small anisotropic dielectric particle ," {\itshape Optics Express} \textbf{18}, pp. 3556-3567 (2010) [doi: 10.1364/OE.18.003556].
 \bibitem{Palik} E. D. Palik, R. Kaplan, R. W. Gammon, H. Kaplan, R. F. Wallis, and J. J. Quinn, "Coupled surface magnetoplasmon-optic-phonon polariton modes on InSb," {\itshape Phys. Rev. B} {\bf 13}, 2497 (1976) [doi: 10.1103/PhysRevB.13.2497].
  \bibitem{BohrenHuffman} C. F. Bohren and D. R. Huffman, {\itshape Absorption and Scattering of Light by Small Particles,} Wiley \& Sons, New York, (1998).
 \bibitem{Bliokh1} K. Y. Bliokh and F. Nori, "Transverse and longitudinal angular momenta of light," {\itshape Phys. Rep.} {\bf 592}, 1 (2015) [doi: 10.1016/j.physrep.2015.06.003].
 \bibitem{Bliokh} K. Y. Bliokh, A. Y. Bekshaev,  and F. Nori, "Optical Momentum, Spin, and Angular Momentum in Dispersive Media," {\itshape Phys. Rev. Lett.} {\bf 119}, 073901 (2017) [doi: 10.1103/PhysRevLett.119.073901].
  \bibitem{Piesbergen1963} U. Piesbergen, "Die durchschnittlichen Atomw\"{a}rmen der A$^{III}$B$^{V}$-Halbleiter AlSb, GaAs, GaSb, InP, InAs, InSb und die Atomw\"{a}rme des Elements Germanium zwischen 12 und 273$^\circ$ K," {\itshape Z. Naturforschung} {\bf 18a}, pp. 141-147 (1963).
\end{thebibliography}
\end{document}